\documentclass[11pt]{article}
\usepackage{amsmath,geometry,amssymb,graphicx,authblk,fancyhdr,algorithm,algpseudocode,mathtools,enumerate,listings,xcolor,url,booktabs,multirow,array,subfiles,adjustbox}

\definecolor{codegreen}{rgb}{0,0.6,0}
\definecolor{codegray}{rgb}{0.5,0.5,0.5}
\definecolor{codepurple}{rgb}{0.58,0,0.82}
\definecolor{backcolour}{rgb}{0.95,0.95,0.92}

\lstdefinestyle{mystyle}{
  backgroundcolor=\color{backcolour},   commentstyle=\color{codegreen},
  keywordstyle=\color{magenta},
  numberstyle=\tiny\color{codegray},
  stringstyle=\color{codepurple},
  basicstyle=\footnotesize\ttfamily,
  breakatwhitespace=false,         
  breaklines=true,                 
  captionpos=b,                    
  keepspaces=true,                 
  numbers=left,                    
  numbersep=5pt,                  
  showspaces=false,                
  showstringspaces=false,
  showtabs=false,                  
  tabsize=2
}

\usepackage{listings, xcolor}

\definecolor{verylightgray}{rgb}{.97,.97,.97}

\lstdefinelanguage{Solidity}{
	keywords=[1]{anonymous, assembly, assert, balance, break, call, callcode, case, catch, class, constant, continue, contract, debugger, default, delegatecall, delete, do, else, emit, event, export, external, false, finally, for, function, gas, if, implements, import, in, indexed, instanceof, interface, internal, is, length, library, log0, log1, log2, log3, log4, memory, modifier, new, payable, pragma, private, protected, public, pure, push, require, return, returns, revert, selfdestruct, send, storage, struct, suicide, super, switch, then, this, throw, transfer, true, try, typeof, using, value, view, while, with, addmod, ecrecover, keccak256, mulmod, ripemd160, sha256, sha3}, 
	keywordstyle=[1]\color{blue}\bfseries,
	keywords=[2]{address, bool, byte, bytes, bytes1, bytes2, bytes3, bytes4, bytes5, bytes6, bytes7, bytes8, bytes9, bytes10, bytes11, bytes12, bytes13, bytes14, bytes15, bytes16, bytes17, bytes18, bytes19, bytes20, bytes21, bytes22, bytes23, bytes24, bytes25, bytes26, bytes27, bytes28, bytes29, bytes30, bytes31, bytes32, enum, int, int8, int16, int24, int32, int40, int48, int56, int64, int72, int80, int88, int96, int104, int112, int120, int128, int136, int144, int152, int160, int168, int176, int184, int192, int200, int208, int216, int224, int232, int240, int248, int256, mapping, string, uint, uint8, uint16, uint24, uint32, uint40, uint48, uint56, uint64, uint72, uint80, uint88, uint96, uint104, uint112, uint120, uint128, uint136, uint144, uint152, uint160, uint168, uint176, uint184, uint192, uint200, uint208, uint216, uint224, uint232, uint240, uint248, uint256, var, void, ether, finney, szabo, wei, days, hours, minutes, seconds, weeks, years},	
	keywordstyle=[2]\color{teal}\bfseries,
	keywords=[3]{block, blockhash, coinbase, difficulty, gaslimit, number, timestamp, msg, data, gas, sender, sig, value, now, tx, gasprice, origin},	
	keywordstyle=[3]\color{violet}\bfseries,
	identifierstyle=\color{black},
	sensitive=false,
	comment=[l]{//},
	morecomment=[s]{/*}{*/},
	commentstyle=\color{gray}\ttfamily,
	stringstyle=\color{red}\ttfamily,
	morestring=[b]',
	morestring=[b]"
}

\algnewcommand\algorithmicinput{\textbf{Input:}}
\algnewcommand\algorithmicoutput{\textbf{Output:}}
\algnewcommand\algorithmicinitialize{\textbf{Initialize:}}
\algnewcommand\algorithmicadd{\textbf{Add:}}
\algnewcommand\algorithmicmultiply{\textbf{Multiply:}}
\algnewcommand\algorithmicmethod{\textbf{Method:}}
\algnewcommand\algorithmicrandnum{\textbf{Random Values:}}
\algnewcommand\algorithmictriple{\textbf{Triples:}}
\algnewcommand\algorithmicdefine{\textbf{Define:}}
\algnewcommand\algorithmickeygen{\textbf{KeyGen:}}
\algnewcommand\algorithmicprover{\textbf{Prover:}}
\algnewcommand\algorithmicverifier{\textbf{Verifier:}}

\algnewcommand\Input{\item[\algorithmicinput]}
\algnewcommand\Output{\item[\algorithmicoutput]}
\algnewcommand\Initialize{\item[\algorithmicinitialize]}
\algnewcommand\Add{\item[\algorithmicadd]}
\algnewcommand\Multiply{\item[\algorithmicmultiply]}
\algnewcommand\Method{\item[\algorithmicmethod]}
\algnewcommand\RandomValues{\item[\algorithmicrandnum]}
\algnewcommand\Triples{\item[\algorithmictriple]}
\algnewcommand\Define{\item[\algorithmicdefine]}
\algnewcommand\KeyGen{\item[\algorithmickeygen]}
\algnewcommand\Prover{\item[\algorithmicprover]}
\algnewcommand\Verifier{\item[\algorithmicverifier]}

\begin{document}
\title{ARPA Whitepaper v0.7}
\author[*]{Derek Zhang}
\author[*]{Alex Su}
\author[*]{Felix Xu}
\author[*]{Jiang Chen}
\affil[*]{Arpachain\authorcr{about@arpachain.io}}

\date{}
\maketitle
\thispagestyle{empty}

\thispagestyle{empty}
\begin{abstract}

\noindent We propose a secure computation solution for blockchain networks. The correctness of computation is verifiable even under malicious majority condition using information-theoretic Message Authentication Code (MAC), and the privacy is preserved using Secret-Sharing. With state-of-the-art multiparty computation protocol and a layer2 solution, our privacy-preserving computation guarantees data security on blockchain, cryptographically, while reducing the heavy-lifting computation job to a few nodes. This breakthrough has several implications on the future of decentralized networks. First, secure computation can be used to support Private Smart Contracts, where consensus is reached without exposing the information in the public contract. Second, it enables data to be shared and used in trustless network, without disclosing the raw data during data-at-use, where data ownership and data usage is safely separated. Last but not least, computation and verification processes are separated, which can be perceived as computational sharding, this effectively makes the transaction processing speed linear to the number of participating nodes.

Our objective is to deploy our secure computation network as an layer2 solution to any blockchain system. Smart Contracts\cite{smartcontract} will be used as bridge to link the blockchain and computation networks. Additionally, they will be used as verifier to ensure that outsourced computation is completed correctly. In order to achieve this, we first develop a general MPC network with advanced features, such as: 1) Secure Computation, 2) Off-chain Computation, 3) Verifiable Computation, and 4)Support dApps' needs like privacy-preserving data exchange.

The remainder of this paper is organized as follows: Section \ref{sec:motivations} introduces the real world motivations which inspired us to build a secure computation network. Following motivations, we highlight our contributions in section \ref{sec:contributions}. We then cover the background of secure computation, along with a comparison of similar technologies.  Our system overview is presented in section \ref{sec:sys_overview}. There, we briefly describe our system design and implementation. In section \ref{sec:mpc}-\ref{sec:crypto}, we discuss, in detail, the major components of our multiparty computation protocol, secure computation process, and considerations in cryptoeconomics. Lastly, we review the implications and applications of the real world; this includes ecosystem design, business cases, and milestones.
\end{abstract}

\newpage
\setcounter{page}{1}
\tableofcontents

\newpage
\setcounter{page}{1}
\section{Motivations} \label{sec:motivations}

\subsection{Off-Chain Computation}
Current blockchain technology reaches on-chain consensus in such a way that a smart contract \cite{smartcontract} runs on every blockchain node. There is no final authority in a decentralized network, such as Bitcoin \cite{nakamoto2008bitcoin}, so that every miner has to validate each transaction before accepting and recording it on the blockchain. In other words, there is no way to verify the result without actually running it on a trustless network. The most prominent problem is the amount of transaction fee for every transaction. As the network grows, the amount of computation in the network combined can easily exceed the gas limit\cite{gaslimit}, causing more forks and raising more security issues.

Off-chain Computation solves this problem by bringing the computation work off the public network, and by verifying the result in public, only when disputes arise. The key in off-chain computation is to have verifiable property in the out-sourced computation task. This property effectively solves the scalability issue\cite{scalability_issue} on blockchain networks. Therefore, we can achieve the throughput of the network being linear to the participating notes.

\subsection{Secure Computation}
One of the ostensible selling points of blockchain technology is its potential to bring greater transparency to financial markets. At its core, blockchain is a means for humans to conduct secure, verifiable, and recordable transactions online without a centralized party. As consensus is reached in public and the result is auditable, the blockchain is claimed to be transparent and public-accessible by design.

However, full transparency can be problematic in the real world. Imagine you opened an account at a bank and soon find out that its ledgers are public, where anyone can access the transaction history of your account (and others). Or, if you made a transaction using Bitcoin at an ice-cream truck and the cashier would be able to know how much money is in your account.

Unfortunately, a privacy feature is not shipped with today's blockchain technology. The growing demand for blockchain and smart contract \cite{sc1997idea,bartoletti2017empirical,ethereum2018} technologies sets the challenge to protect users from intellectual property theft and other attacks\cite{atzei2017survey}: security, confidentiality and privacy are the key issues holding back the adoption of blockchain\cite{banker2016privacy}.

ARPA is built to provide secure computation capability as an off-chain solution for most blockchain networks.

\subsection{Centralized Dataset}
Large internet companies collect data from users' online activities to train recommendation systems that predict the customers' future interest and actions. Health data from different hospitals and government organizations can be used to produce new diagnostic models, while financial companies and payment networks can combine transaction history, merchant data, and account holder information to train more accurate fraud-detection engines.

All this data needs to be aggregated from an individual level, but the profits realized from the data have never been shared with the data contributors. The General Data Protection Regulation (GDPR\cite{GDPR2018}) made it clear that data holders have the responsibility to facilitate the portability of the data. To really enforce this regulation, our data should not be stored at the company’s hard drive as it faces the inevitable risk of being breached. 

Unfortunately, companies have found it difficult to protect their critical data from determined attackers. The convention with data holding is ill-formed in the following ways:

\begin{enumerate}
\item Centralized data-at-risk
\item Unfair value distribution - individual users never get paid for their data
\item Established player dominance creates an unfriendly moat that deters new and innovative challengers from competing.
\item Talented individuals can't build a good model without joining the company
\end{enumerate}

We need a decentralized data sharing scheme that enable everyone to build models on it, reward those who contribute to it, and at the same time keep the data secure.

\subsection{Data Security in Exchange}
\begin{figure} [ht]
\centering
\includegraphics[width=6cm]{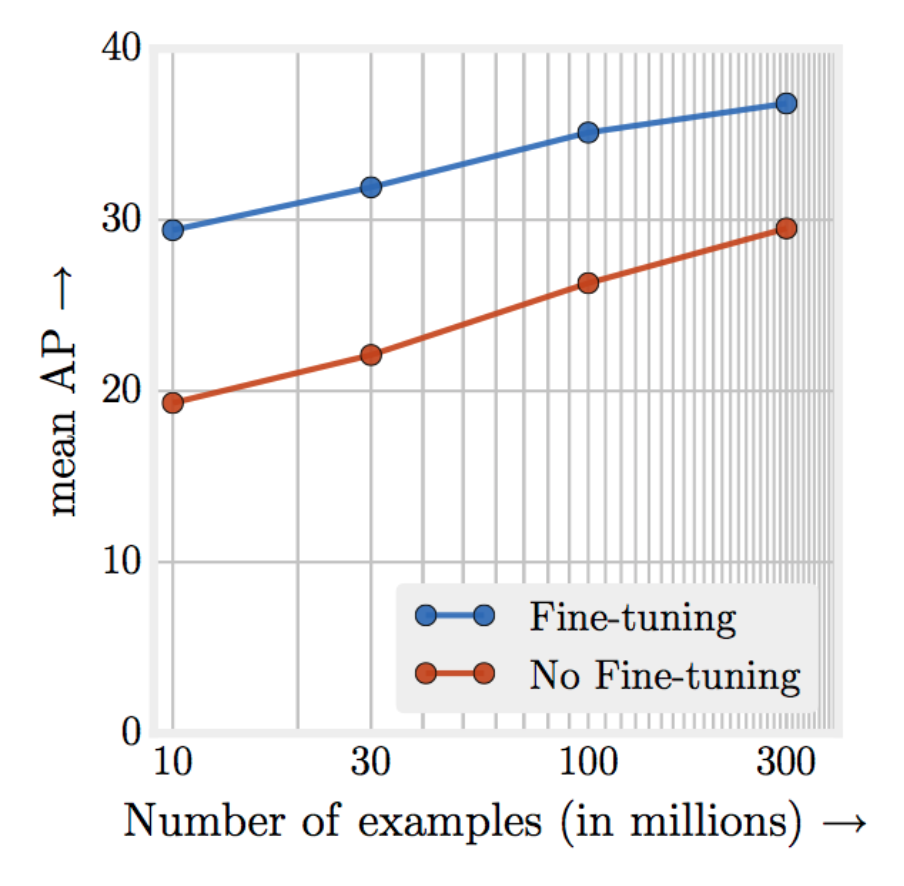}
\caption{Data size and model performance}
\label{fig:model_perf}
\end{figure}

There is a tension that arises when individuals, companies, or governments deal with sensitive data. On the one hand, data science is a fundamental component of the informational age. We often hear that “data is the new oil”: there is immense financial and social value in acquiring raw data. A recent paper from Google \cite{sun2017revisit} (Figure \ref{fig:model_perf}) confirms a well-known fact that data size is positively correlated to model performance, regardless of model quality. It explains the motivation for companies to collecting extensive amount of user data.

On the other hand, the more value there is to be gained by piecing data together, the more cautious everyone has become about data sharing, since data breaches can cause financial, legal, and political harms. 
This appears to be a central trade-off: we can share data in order to learn new insights that benefit the society as a whole, or we can isolate data into protected silos that safeguard our individual privacy. 

Privacy-preserving computation is the solution for this paradox. 

\section{Contributions}  \label{sec:contributions}

\paragraph{Secure Computation} Computation is carried out securely so that no participating node can learn anything more than its prescribed output.

\paragraph{Verifiable Computation} Computation can be audited publicly, and its correctness can be proven. Therefore, it is possible to outsource computation from the blockchain network.

\paragraph{Layer2 Solution} Combining secure and verifiable computation, the heavy-lifting work of computation is done off-chain. Essentially, making our secure computation protocol adaptable to any existing blockchain network.

\paragraph{Scalability} ARPA is designed as a Layer2 solution. Because the verification has complexity of O(1), the on-chain network will never reach its computation (gas) limit. Therefore, we can improve the computation scalability and TPS(transaction per second) of any network. The computation capacity is increased linearly to participating nodes.

\paragraph{Efficiency} State-of-the-art implementation of MPC protocol is used to speed-up the secure computation.Though this implementation, 5-6 magnitudes of speed improvement are achieved compared to Fully Homomorphic Encryption (FHE).

\paragraph{Availability} World's first general purpose MPC network for secure computation. With high availability and low cost, we promote data security and privacy practice that is difficult to achieve otherwise.

\section{Background} \label{sec:background}
\subsection{Multiparty Computation (MPC)}

Multiparty Computation(MPC) is a way by which multiple parties can compute some function of their individual secret inputs, without any party revealing anything to the other parties about their input, other than what can be learned from the output. Unlike traditional cryptographic tasks, where the adversary is outside the system of sender and receiver, the adversary is a part of the participants. This model frees the cryptographers from the centralized paradigm to the distributed paradigm. Many unsolved problems like Yao's millionaires' problem \cite{yao1982protocols} or secure auction system \cite{bogetoft2009secure} can be perfectly settled. In a sense, MPC is rather a new paradigm than a specific cryptographic algorithm.

MPC can be considered as a configurable framework with a mixture of crypto tools, which gives participants the right to organize the specific protocol according to the security level and efficiency limitation of a scenario. The beauty of multiparty protocols is that they use a rich body of tools and sub-protocols, some of which have been developed especially for MPC and others previously developed for the cryptographic non-distributed setting. These tools include zero-knowledge proof (ZKP) \cite{rackoff1991non}, probabilistic encryption, information-theoretic Message Authentication Code (MAC), various distributed commitment schemes, and oblivious transfer \cite{rabin2005exchange}. Most importantly, secret-sharing \cite{pedersen1991non} and computing with shares of a secret is fundamental to achieving secure multiparty computation. In particular, the polynomial secret sharing of Shamir \cite{shamir1979share} in the case of passive adversary is a cornerstone in multiparty computations, and the verifiable secret sharing \cite{chor1985verifiable} plays an analogous role in the Byzantine adversary case. 

There are various MPC protocols built with different assumptions in terms of security models. For example, Sharemind \cite{bogdanov2008sharemind} and VIFF \cite{geisler2007viff} assume a semi-honest scenario where the adversary will follow the given protocol but only try to reveal others' private input. Whereas, under malicious situation the participants will manipulate whatever they need to get the data, including colluding and/or malicious computation. 

\begin{figure}[ht]
\centering
\includegraphics[width=10cm]{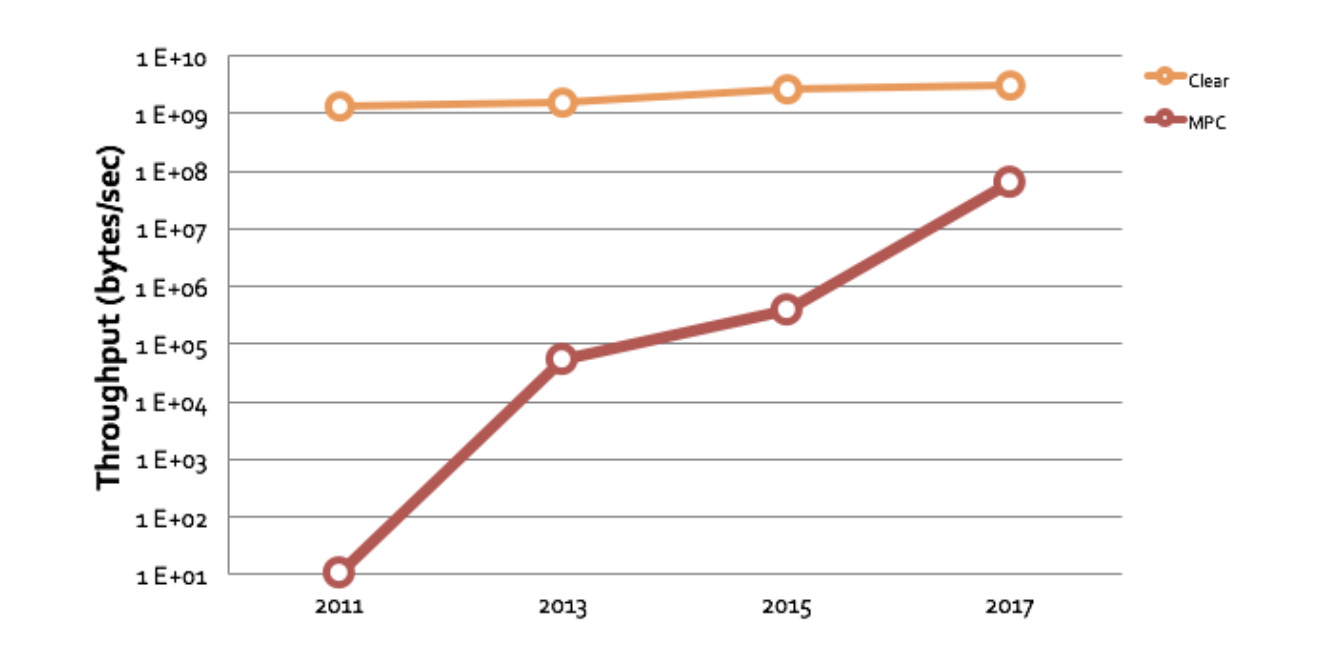}
\caption{Efficiency Improvement over MPC}
\label{fig:mpc_efficiency}
\end{figure}

With theoretical and practical achievements over the last decade, MPC has finally evolved to a point where performance shouldn't be considered the primary impediment to use. With theoretical constructions going back 35 years, there are substantial improvements in algorithmic and engineering designs over the past decade to improve performance. The performance of MPC has improved by 4-5 orders of magnitude over the past decade (Figure \ref{fig:mpc_efficiency}). Both for comparison purposes and to account for the effects of Moore's law, we also show the performance of native AES computation over the same time period.

Our goal is to build a MPC network with high availability for the first time, where any business needs for secure computation can be conducted on the network or by using smart contracts on existing blockchains such as Ethereum or EOS. This will bring several benefits: 1) MPC computation will be as simple as plug-and-play. No prior knowledge is needed to set it up correctly. 2) Since dedicated nodes run MPC, the cost for conducting such work will be much lower. and 3) The MPC "cloud" can bring awareness to the business world, as well as the general public, so that more secure computation can be conducted using MPC.

\subsection{Other Candidates of Secure Computation}
In the past ten years, several technologies towards practical secure computation protocols have been proposed and studied. These efforts have been classified into three primary domains, i.e. homomorphic encryption (HE), multiparty computation (MPC), and trusted execution environment (TEE). 

\subsubsection{Homomorphic encryption}
Homomorphic Encryption(HE) is a form of encryption that allows computation on ciphertext, generating an encrypted result which, when decrypted, matches the result of the operations as if they had been performed on the plaintext. With such a tool, one could outsource storage and/or computation without endangering data privacy. Because HE allows calculation on encrypted data while remaining encrypted, it has been extensively researched as a candidate for secure computation. 
\begin{figure}[ht]
\centering
\includegraphics[width=10cm]{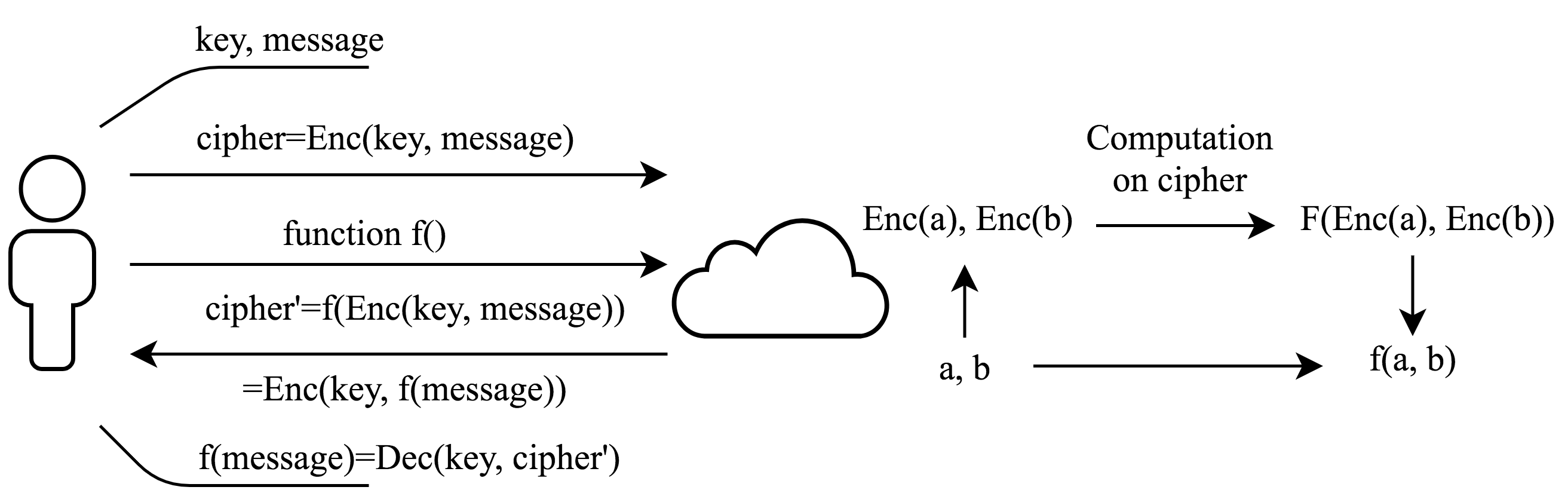}
\caption{Homomorphic Encryption}
\label{fig:HE}
\end{figure}

The first concrete HE scheme was proposed in 2009, according to the work of Gentry. It introduced an interesting structure, as well as, a nice trick called bootstrapping to reduced the inherent noise that accompanied the running of additions and multiplications. This allowed practical somewhat homomorphic encryption (SHE) to develop into fully homomorphic encryption (FHE).

Nevertheless, cutting-edge homomorphic schemes cannot provide an efficient way to compute large depth arithmetic circuits due to the following constraints. First, ``bootstrapping'' adds an extra cost to an already quite heavy process. Currently, practical use of HE focuses mainly on optimization of an evaluated function, which avoids expensive processes by limiting the circuit multiplication depth. In addition, using HE schemes will lead to a huge ciphertext expansion, by the overhead from 2,000 to 500,000 or even 1,000,000 (times) according to the scheme and the targeted security level. This is due to the fact that homomorphic schemes must be probabilistic to ensure semantic security, and to the particular underlying mathematical structures. As we can see, SHE schemes are the most promising today in HE variants, and it will be utilized in our secure computation program mentioned later.

\subsubsection{Zero-Knowledge Proof (ZKP)}
Zero-Knowledge Proof (ZKP) is a method by which one party (the prover Peggy) can prove to another party (the verifier Victor) that she knows a value x, without conveying any information apart from the fact that she knows the value x. Recent blockchain project was developed to leverage ZKP as a trusted off-chain computation solution. In this protocol, the function is compiled into a circuit and transmitted to a third-party execution environment where the data will be evaluated using the circuit. Similar to FHE scheme, it cannot prove that the actual amount of work being done in remote environment. In addition to that, ZKP simply cannot guarantee the computation is secured from hacker of malicious party.

\subsubsection{zk-SNARK}

Zero-Knowledge Succinct Non-Interactive Argument of Knowledge (zk-SNARK) is a protocol which creates a framework in which a person -- called prover -- can quickly convince another person -- called verifier -- that she or he ``knows'' a secret without revealing anything about the secret. The first constructions of SNARK protocols were inspired by the Probabilistically Checkable Proof (PCP) theorem which shows that (Nondeterministic Polynomial time) NP problems statements have ``short'' PCPs\cite{micali2000computationally}. New instantiations were found which allow faster and shorter proofs, when a pre-processing state is permitted.

zk-SNARKs intend to enhance the privacy of users transacting on the Zcash blockchain. With cryptocurrencies such as Bitcoin, an individual can identify user addresses and track the movement of value between transacting parties on the blockchain. In this case, Bitcoin only provides users with pseudonymous protection, rather than complete anonymity. zk-SNARKs are designed to solve this problem by completely encrypting user transaction information on the Zcash blockchain. An abstract zk-SNARK description can be denoted as Alg. \ref{alg:zksnark}

\begin{algorithm}[htbp]
\caption{zk-SNARK}
\label{alg:zksnark}
\begin{algorithmic}[1]
\KeyGen $(vk,pk)\gets \Call{KeyGen}{\text{circuit } C,\lambda}$.
\Prover $\pi \gets \Call{Prover}{pk, \text{public input } x, \text{secret input } w}$
\Verifier $\exists w$ s.t. $C(w, x) \gets \Call{Verifier}{\pi,vk,x}$
\end{algorithmic}
\end{algorithm}

\subsubsection{Trusted Execution Environment (TEE)}
Trusted Execution Environment (TEE) is a tamper resistant processing environment that runs on a separation kernel [2]. It is another trend to solve a secure computation problem. An ideal TEE guarantees the authenticity of the executed code, the integrity of the runtime states (e.g. CPU registers, memory and sensitive I/O), and the confidentiality of its code, data and runtime states stored on a persistent memory. In addition, it shall be able to provide remote attestation that proves its trustworthiness for third-parties.

Hardware manufacturers are eager to propose their own trusted hardware solutions but lack a general standard over different platforms. Most prominent process unit designers have embedded their hardware secure module in their products (Ex. Intel Software Guard Extensions (SGX), ARM TrustZone, AMD Secure Encrypted Virtualization (SEV) and NVIDIA Trusted Little Kernel (TLK). 
\begin{figure}[ht]
\centering
\includegraphics[width=10cm]{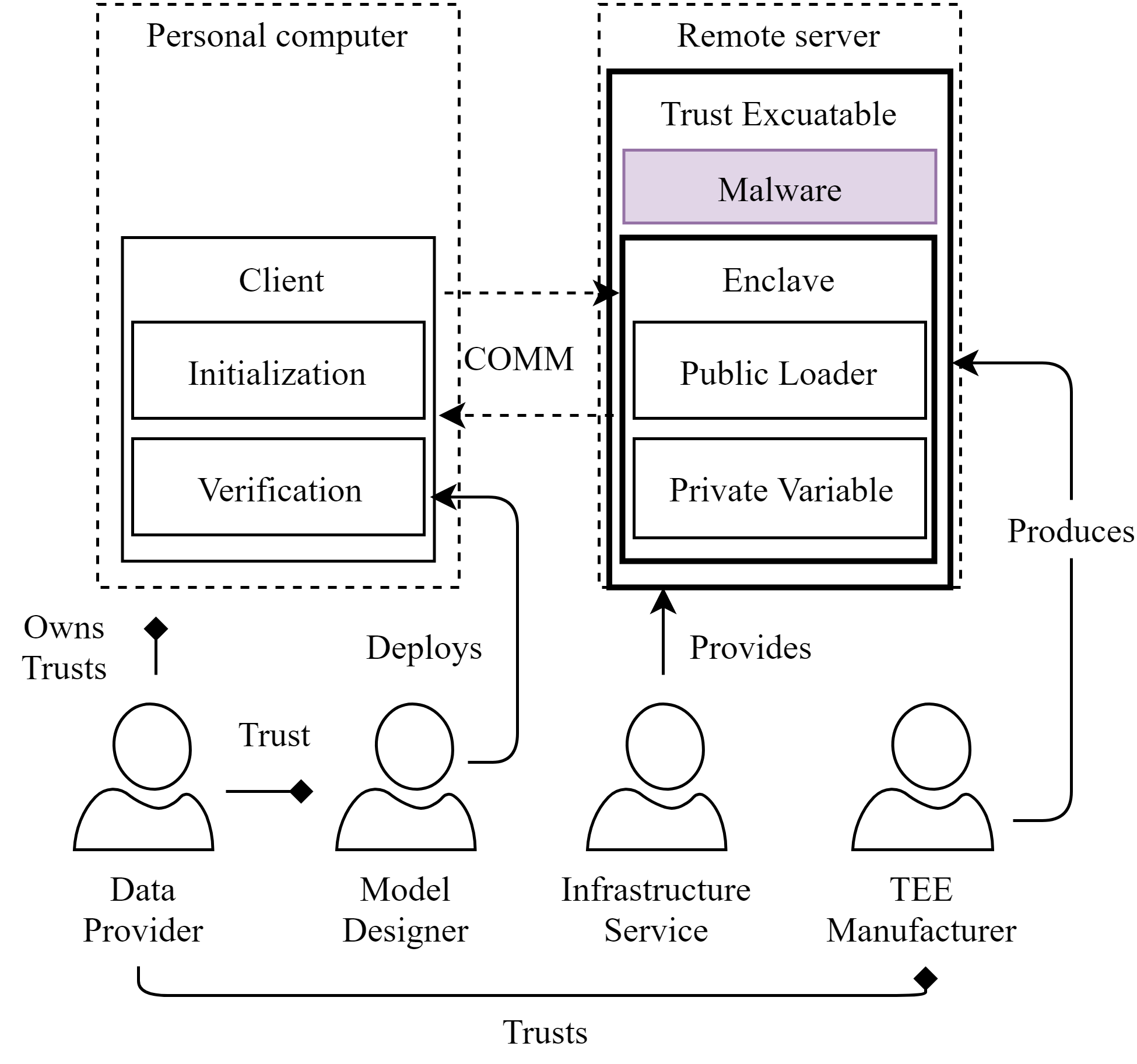}
\caption{Trust Execution Environment}
\label{fig:TEE}
\end{figure}

Software Guarded Execution (SGX) is a set of instructions promoted by Intel to enable code to be run at private region of CPU. The benefit of such implementation is that the application is secured at hardware level with cost of developing a software. Most solutions to provide privacy to smart contracts are based on SGX, which was a sensible choice years ago. Several projects existed are using SGX as off-chain computation solution. 

However, recent attacks have proven SGX to be inadequate. This seemingly secure protocol is, in fact, not secure at all. The remote attestation does not prevent a malicious cloud service provider from first faithfully responding to remote attestation queries, but then emulate the rest of the protocol (such as KeyGen and CSR) outside of the enclave. In other words, SGX is not a protocol designed for ``Universal Composition'' (UC) where, the real-world behavior and the ideal-world definition (function) of a protocol are computationally indistinguishable for every adversary-controlled environment. In simple term, with TEE, one can trust the hardware but not the person controlling the hardware. Therefore, the SGX stack is best used in a permissioned network\cite{permissioned}, where all nodes are pre-approved and the environment is certified and trusted.

Furthermore, SGX instructions can only be used on Intel's CPU. In cases like deep learning, where algorithms are accelerated on GPU, or in situation that Intel is not considered to be trustworthy, the SGX solution will bear significant risk. In addition, the proper use of TEE is also a nontrivial work, especially the collaboration of different TEE modules.

\subsection{Comparison} \label{sec:comparison}

\subsubsection{Trust Dependency}
Privacy-preserving computation can be sorted into two development directions: \textit{secure computation} and \textit{trusted computation}. The prior mainly focuses on the cryptography security brought by hard problems that are elaborately designed using mathematics. The security is derived purely from the number theory, which means even with unaffordable computing power, an adversary cannot break the encryption. The latter one starts from a different point, which tries to enforce the computing involvement to consistently behave in the expected way. Enforcing the behavior is achieved by hardware authentication, memory encryption, specific instruction design, etc. These technologies are always bound to the hardware platform such as the memory storage or processing elements.

Obviously, both Homomorphic Encryption and MPC belong to secure computation, and Trusted Execution Environment(TEE) is the effort of trusted computation. In brief, secure computation is more suitable to be used in permissionless network while trusted computation relies more on physical entities and can be used in a permissioned network such as Hyperledger\cite{turkihoneyledgerbft}.

\subsubsection{Scalability and Flexibility}
Given the foundation of mentioned computation methods, it can be observed that one can evaluate a secure computation on any computing power, either a data center or an edge device such as an automobile and phone. Due to the TEE's design methodology, users have to transplant the program from one platform to another by fitting their work into the safes. This feature makes the TEE less flexible than the pure protocol solutions. Considering the underlying device independency, MPC or HE can take advantage of mutual tool chains of hardware manufacturers and have future potentials for hardware accelerations.

As for different scenarios of computation, MPC can transform into a suitable pattern to make a trade-off between computation complexity and secureness. The nature of MPC can also satisfy various data-privacy input conditions. HE may support multiparty input as well by using distributed encryption and decryption, but the overhead of bootstrapping limits the secure functions evaluated on the HE.

\subsubsection{Practical Efficiency}

Thanks to the continuous effort of cryptographers and computer scientists in the last decade, different frameworks of MPC designed to solve various type of cases have been invented. We will develop a general purpose MPC network that achieves security in a malicious, dishonest, majority n-party setup.

Taking Intel SGX\cite{sgx} as the representative of TEE, the main overhead of SGX, compared with plaintext manipulating, is caused by fetching input from normal memory to the enclave local memory that involves encryption. According to the experiments on the enclaves, the Intel enclave instructions did not achieve the best performance. On the other hand, the optimal algorithm on plaintext cannot get the same throughput when processing enclave local memory. 

For S/FHE and MPC implementation, multiplication over ciphertext is the most frequently called and fundamental operation. Table \ref{tab:HE_vs_MPC} lists the state-of-the-art literal performance for both systems. The final column shows the latency of multiplication for different parameter settings. Although the diagram compares hardware implementation of SHE\cite{bonnoron2017somewhat} with practical software MPC\cite{keller2018overdrive}, it can be observed that MPC is 1-2 orders of magnitude faster than SHE. Considering in the future, dedicated hardware acceleration of MPC building blocks will be developed, and there should be another 10-100 times throughput improvement potential for MPC. Therefore, there exists 3-4 orders of magnitude of performance gap between HE and MPC. Furthermore, the most expensive operation of FHE is bootstrapping which enables unlimited multiplication depth is not listed in table. This may bring another slowdown in FHE computation.

\begin{table}[ht]
\centering
\begin{tabular}{*5c}   \toprule
Category     &   Scheme   &   \multicolumn{2}{c}{Security parameter} &       Operation   \\  \midrule
\multirow{5}{*}{HE} & \multicolumn{1}{m{3cm}}{Hardware implementation}   &    n   &   q     &   homomorphic multiplication \\  \cmidrule{2-5}
                    &   YASHE-NTT               &  4096  &   125 bits  &    6.5ms   \\
                    &   YASHE-NTT               &  16384(SIMD)  &  512 bits   &    48ms    \\
                    &   YASHE-NTT               &  32768(SIMD)  &  1228 bits  &    121ms    \\ \midrule
\multirow{4}{*}{MPC} &  Software implementation &  \multicolumn{2}{c}{Statistical security parameter}  &   Mult triples generations  \\ \cmidrule{2-5}
                     &  MPC                    &  \multicolumn{2}{c}{128 bits}                     &      0.23ms   \\
                     &  Overdrive: High Gear    &  \multicolumn{2}{c}{128 bits}                     &      0.43ms   \\
                     &  Overdrive: Low Gear     &  \multicolumn{2}{c}{128 bits}                     &      0.067ms  \\ \bottomrule
\end{tabular}
\caption{Performance Comparison of HE and MPC}
\label{tab:HE_vs_MPC}
\end{table}

\subsubsection{Conclusion}
In this section we discussed the history of MPC and its cryptographic properties and demonstrated that it is actually a very good candidate for mutually-distrusted parties to conduct secure computation together – a very similar setting to the blockchain network. 

We also visited several technologies currently existing as candidates of secure computation, namely Homomorphic Encryption, Zero-Knowledge Proof, and Trusted Execution Environment. Our discoveries indicate that, while some technologies have advantage such as computation efficiency, they are not offering security and features needed in a permissionless network. In essence, we need to be able to verify the secureness, correctness and privacy-preservingness of computation. Detailed aspects under consideration are as follows:

\paragraph{Efficiency} The speed of computation. Our experiments have found that, TEE is on par with clear text computation. Followed by MPC, about one magnitude slower, as forementioned. As for ZK-SNARK, it has around $10^3$ times overhead\cite{ben2013snarks}. The least favorable choice in terms of efficiency is FHE, with about $10^7$ to $10^8$ times slower.\cite{bonnoron2017somewhat}

\paragraph{Privacy-preserving} Privacy-preserving here refers to the capability of evaluating a function on a dataset, while not revealing the detail to anyone. This is the core of secure computation.

\paragraph{Proof of Correctness} Prove that the computation work is actually using the prescribed function.

In a trustless network, it is very important to prove that a certain function is evaluated in a correct way. In blockchain, it is simply done by repeated computation on every node running the smart contract. Consensus can be made once every participating node reaches on the same result. However, in the context of secure computation, computation is not carried out on every node but delegated to certain node(s) for the task. One needs to submit proof that computation has done the work with prescribed routine, or the result cannot be trusted. 

\paragraph{Proof of Computation} Prove the amount of work the participating node has done.

In Bitcoin and other PoW-style blockchain network, participating nodes prove their work by solving a hard problem (puzzle) and submit the hash value according to the nonce determined at the beginning of the block. It is trivial work for nodes to verify the result (hash) and prove that the node has indeed solved the puzzle. In a computation network, it is also important to know how much work has been done in the computation, so the node can be rewarded with a corresponding token. In MPC, the amount of computation can be proved by counting the triples consumed during the computation.

\paragraph{Proof of Secureness} Prove that the computation is actually carried out in the secured environment.

When computation is not repeated on every node, one needs to submit proof that the computation is carried out in a secured environment. Although this is what TEE is designed for, this seemingly naive question is actually very hard to prove. As mentioned earlier, secure computation is a cryptography process that guarantees the computation process is following the protocol or the result cannot be accepted. But trusted computation, on the other hand, enforces the process to be consistent with designed security (i.e. run the function) using trusted hardware. Nevertheless, in a permissionless network, one cannot convince other nodes to trust his result just by claiming it got remote attestation from a centralized server. Furthermore, even the remote attestation proves the security of the environment, one cannot prove the code is executed in that trusted environment.

A final conclusion with all facts related to blockchain construction is listed in table \ref{fig:secure_computation_comparison}.

\begin{table}[ht]
\centering
\begin{tabular}{*5c}   \toprule
        &	MPC	&	FHE	&	ZK-SNARK	&	TEE\\ \midrule
Efficiency	&	Acceptable to Fast	&   Prohibitively Slow	&	Acceptable	&	Fast	\\
Privacy-preserving	&	Yes	& Yes	&	No	&	No 	\\
Trustless	&	Yes 	&	Yes	& 	No 	&	No	\\
Proof of Correctness	&	Yes	&   No     &	Yes	&   No	\\
Proof of Computation	&	Yes	&	No	&	No 	&   No	\\
Proof of Secureness	&	Yes	&	Yes	&	No 	&	No	\\ \bottomrule
\end{tabular}
\caption{Secure Computation and Trusted Execution}
\label{fig:secure_computation_comparison}
\end{table}

\newpage
\section{System Overview}  \label{sec:sys_overview}

The ARPA network is designed with two layers: consensus layer and computation layer. The computation layer is responsible for conducting ARPA's Multiparty Computation protocol, while the consensus layer makes sure that transactions and other metadata are recorded and reach consensus. Each blockchain node is comprised of two layers/services.

\begin{figure}[ht]
\centering
\includegraphics[width=12cm]{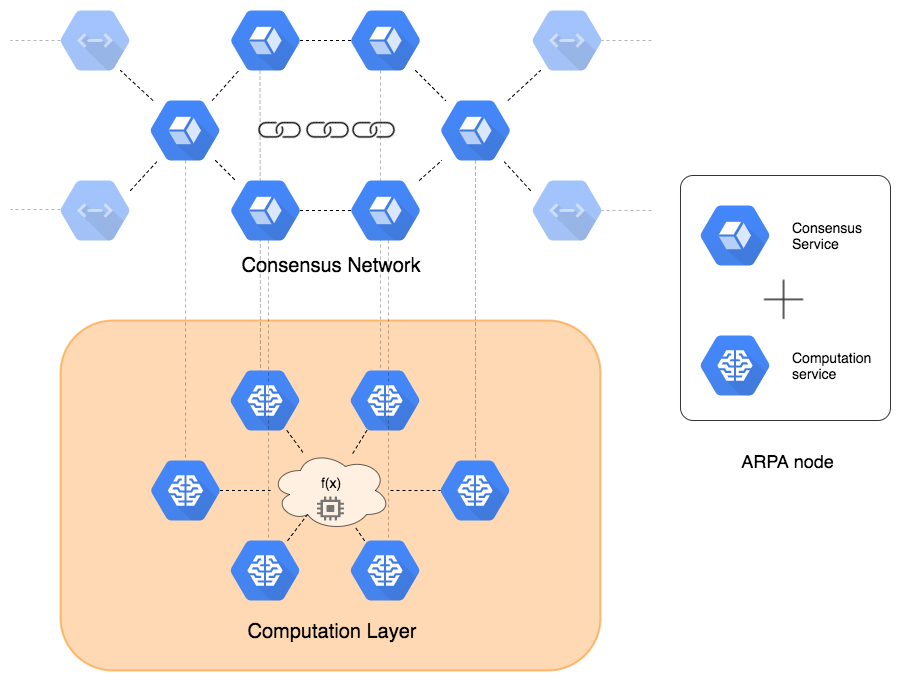}
\caption{ARPA overview}
\label{fig:arpa_overview}
\end{figure}

The system's overall design is illustrated in Figure \ref{fig:arpa_overview}.

\paragraph{Consensus layer} ARPA's network is designed as a permissionless network where participants can join autonomously, contribute their computation resources, and earn economic return. To achieve this, we need to faithfully record the transaction information, as well as verification scheme for the computation. This allows the network to form the quorum based on historical performance (with a credit system). 

The consensus service includes Account Signature, 
Computation Verification, Node Communication, Credit system, and Consensus protocol. 
A decentralized ledger is implemented to record transnational and stateful data. 

We leverage Secret Sharing and Threshold Signature to create decentralized randomness in block generation. This uncertainty is used to prevent anyone from seizing control of the block by creating above average probability of block-signing rights.

\paragraph{Computation layer} This layer is mainly a protocol for participating nodes to jointly perform secure Multiparty Computation (MPC) with arandom quorum, additional coordination, model compilation, Secret Sharing, distributed storage, error handling, computation proof generation, decentralized preprocessing, network load-balancing, and other tools to make sure nodes in a quorum can finish task with designated cryptography features. 

The MPC protocol is designed with an economic incentive/penalty so that it will cost a malicious node more than it can gain from the system.

We will come back to this in Section \ref{sec:crypto}.

\subsection{Interoperability}
The need for privacy-preserving computation is universal. It would be a greater benefit for other popular blockchains, such as Ethereum and EOS, to have secure computation capability. Therefore, APRA's system is designed to be inter-operable with other blockchain systems, while maintaining our own chain as a standalone network for secure computation. To be able to conduct off-chain computation, our network nodes need to monitor smart contract information on other chains, and act upon requests from these smart contracts. This realization of secure computation is called Private Smart Contract. This completely obscures the data from the public, while keeping the computation auditable. Private Smart Contracts will later be described in further detail.

\subsection{ARPA Virtual Machine (AVM)}
Similar to Ethereum Virtual Machine (EVM), The Arpa Virtual Machine is a Turing Complete Virtual Machine that can run smart contracts in it. However, to enable secure computation and other advanced features such as Private Smart Contract, AVM is shipped with a stronger toolset with close to native performance. To this end, AVM adopts WebAssembly(WASM), a binary instruction for a stack-based Virtual Machine, which lets us do more with less. WASM brings faster code, smaller deliverables, and less overhead. The Virtual Machine can execute bytecode compiled from scripting language, such as JavaScript and Python, for better adoption and less learning curve.

To enable the execution of "Private Smart Contract", a smart contract that can protect data privacy, function privacy, participants' identify and their state change, the EVM has a built-in function cryptography library for easier computation dispatch and verification.

AVM also supports native Oracle implementation. External data can be fetched and processed without the need to setup an oracle server. This will be very handy when the computation verification package is sent back from the secure computation layer.

\newpage
\section{ARPA Multiparty Computation Protocol}  \label{sec:mpc}
\begin{figure}[ht]
\centering
\includegraphics[width=12 cm]{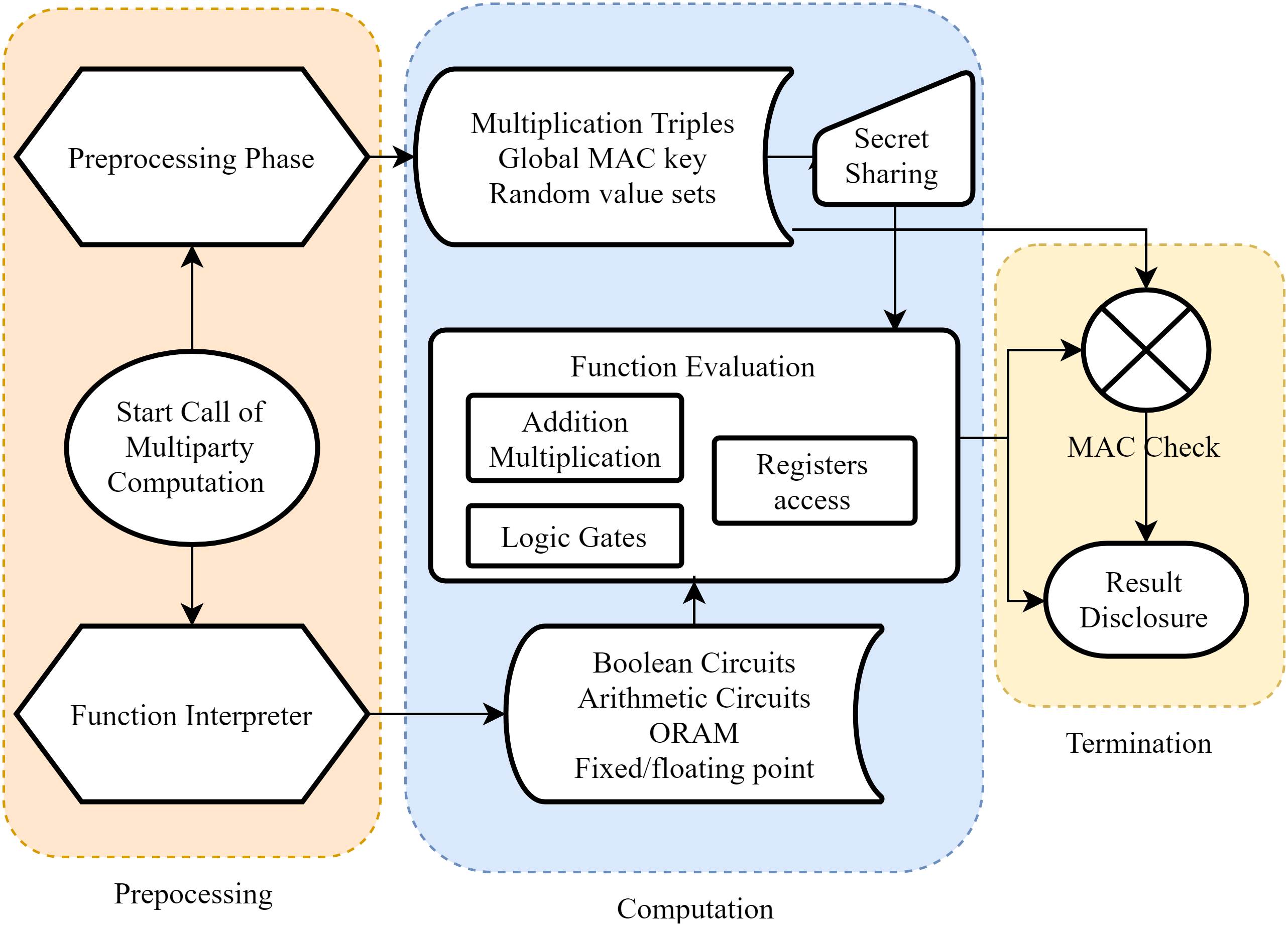}
\caption{ARPA MPC protocol}
\label{fig:mpc_flowchart}
\end{figure}

The protocol implemented in ARPA's computation network is organized as Figure \ref{fig:mpc_flowchart}. The procedure can be viewed as a two-phase protocol: Preprocessing and Computation. The computation phase is the main procedure that involves secret sharing, function evaluation, and revealing outcomes. On the other hand, the preprocessing phase is the foundation which provides necessary raw materials for computation. We separate these two phases apart because preprocessing is independent of both the function and the data.

\subsection{Computation Phase}

During the computation phase, the main task is to securely evaluate an arbitrary circuit on private inputs which is keep secret throughout and reveal the result only when the parties follow the prescribed protocol. This phase can be concluded as a 3-stage protocol: input, evaluation, output. After private inputs are provided and secretly shared, all parties perform addition and multiplication on shared values, collaboratively, according to the arithmetic circuit. Eventually, results are opened after computing verification.

\subsubsection{Sharing secrets}
At the beginning of the computation phase, the input needs to be shared among all parties. This process is called secret sharing. Secret sharing is the method of distributing a secret among a group of participants, each of whom is allocated a share of the secret. The secret can be revealed only when a sufficient number of shares are combined together. Individual shares are unreadable ciphertext on their own.

A secret sharing scheme usually consists of two main functions, $\Call{Share}{}$ and $\Call{Reconstruct}{}$. $\Call{Share}{}$ allows one party to share a secret, $x$, among n parties such that the secret remains secure against an adversary up to $t-1$ parties in a $(t,n)$-threshold scheme. While $\Call{Reconstruct}{}$ allows any group of t or more parties together to open the secrets. In ARPA MPC, we design the secret-sharing scheme to work effectively under universally composable security so that the threshold is set to $n$, meaning our system is secure up to $n-1$ colluded malicious parties.

In the \Call{Input}{} step, parties need to use shares of random values: each party $P_i$ holds some $r_i$ and the secret value $r_i$ is uniformly random and not known by any party. First, every party sends their share $r_i$ of $r$ to party $P_{provider}$. This is called a `partial opening' because now $P_{provider}$ discovers the value of $r$ by summing the shares. Party $P_{provider}$ then broadcasts his input $x-r$. Party $P_1$ sets its share of x as $x_1 = x - r + r_1$, and for $Pi$ sets its x share of as $x_i = r_i$. In practice, it is not always $P_1$ who does a different computation. Now we turn $x$ into $\langle x \rangle$ and shares are $x_i$.

\subsubsection{Evaluating circuits} 

On ARPA's MPC framework, the function to be evaluated should be expressed in an arithmetic circuit representative. A general method to transfer high-level functions into arithmetic circuits are included in our compiler. which is concretely described in the preprocessing phase in section\ref{sec:preprocessing}. This compilation is done prior to computation phase.

The arithmetic circuit basically consists of addition and multiplication operations. In the evaluation step, since the secret sharing scheme is additive and linear, addition can be done via local computation which has no communication cost. Specifically, to compute $\langle a + b\rangle$ or $\langle \alpha \cdot a\rangle$, each party simply does $a_i + b_i$ or $\alpha \cdot a_i$. We then get the $\langle a + b\rangle $ and $\langle \alpha \cdot a\rangle$ in secret shares type.

Multiplication on secret values, however, requires some interaction between the parties. Beaver's triples are used to facilitate the online multiplication. When parties are scheduled to compute $\langle x\cdot y \rangle$ given $\langle x \rangle$ and $\langle y \rangle$ one available `triple' $\langle a \rangle$, $\langle b \rangle$, $\langle c\rangle$, where $c = a \times b$ is taken from the pre-processed raw material. Then, by broadcasting $x_i - a_i$ and $y_i - b_i$, each party can compute $x - a$ and $y - b$. Which now means these intermediates are publicly known. Eventually everyone can locally compute $z_i=(x-a)(y-b)+x_i(y-b)+y_i(x-a)+c_i$, so that they have a secret share of $\langle z \rangle$.

\begin{equation}
\begin{aligned}
\langle z \rangle &= \langle c \rangle + d \cdot \langle b \rangle +e \cdot \langle a \rangle + d \cdot e\\
&=\langle a \cdot b + (x-a) \cdot b+ (y-b) \cdot a + (x-a)\cdot (y-b)\rangle\\
&=\langle x\cdot y \rangle
\end{aligned}
\end{equation}

The upshot is that with enough amount of triples, the MPC engine can perform any multiplication depth computing on secret shared data, and hence we can compute any arithmetic circuit. For the reason that some intermediates like $x-a$ are opened for subsequent procedure, triples cannot be reused because this would reveal information about the secrets.

Importantly, it can be observed that triples are independent of not only the input data, but also the circuit to be evaluated. This means that we can generate these triples at any point prior to evaluating the circuit. The value of triples are not known by any parties when generated. Each party only knows a share of each of `some values' for which they are told this relation holds. Moreover, since addition, scalar multiplication, and field multiplication are inexpensive in terms of communication and computation, the computation phase is both highly efficient and information-theoretically secure. 

With the methodology of addition and multiplication on secret shares, we can build any computable functions on arithmetic circuits. Taking an example of secure signed floating number multiplication like Alg.\ref{alg:fpMul} \cite{aliasgari2013secure} which involves exponents addition, sign bit operation and mantissa multiplication. Specifically, multiplication of two floating point numbers $\langle v_1, p_1, z_1, s_1\rangle$ and $\langle v_2, p_2, z_2, s_2\rangle$ is performed by first multiplying their mantissa $v_1$ and $v_2$ to obtain a $2l$-bit product $v$. The product then needs to be truncated by either $l$ or $l-1$ bits depending on whether the most significant bit of $v$ is $1$ or not. In the protocol below, it is accomplished obliviously on lines 2-4. Partitioning this truncation into two steps allows us to reduce the number of interactive operations at a slight increase in the number of rounds. After computing the zero and sign bits of the product (lines 5-6), we obviously adjust the exponent for non-zero values by the amount of previously performed truncation.

Combining the above, we have now established roughly how to evaluate an arithmetic circuit.

\begin{algorithm}[htbp]
\caption{Secure Floating Number Multiplication}
\label{alg:fpMul}
\begin{algorithmic}[1]
\Require $\langle [v_1],[p_1],[z_1],[s_1]\rangle, \langle [v_2], [p_2], [z_2], [s_2]\rangle$ 
\Ensure $\langle [v], [p], [z], [s]\rangle \gets \Call{FLMul}{\langle [v_1],[p_1],[z_1],[s_1]\rangle, \langle [v_2], [p_2], [z_2], [s_2]\rangle}$
\Method
\State $[v]\gets [v_1][v_2]$;
\State $[v]\gets \Call{Trunc}{[v],2l,l-1}$;
\State $[b]\gets \Call{LT}{[v], 2l,l+1}$;
\State $[v]\gets \Call{Trunc}{2[b][v]+(1-[b][v],l+1,1}$;
\State $[z]\gets \Call{OR}{[z_1],[z_2]}$;
\State $[s]\gets \Call{XOR}{[s_1],[s_2]}$;
\State $[p]\gets ([p_1]+[p_2]+l-[b])(1-[z])$;
\State return $\langle [v], [p], [z], [s]\rangle$;
\end{algorithmic}
\end{algorithm}

\subsubsection{Information Theoretic MAC}

When using multiplication triples and a linear secret-sharing scheme with threshold $t$, we obtain a passively secure MPC protocol that tolerates up to $t$ corruptions. In the dishonest majority setting with up to $n-1$ corruptions, the most natural secret sharing scheme to use is simple additive secret sharing, where $x\in F$ is shared between n parties by distributing n random shares $x_i$ satisfying $x=\sum_1^n x_i $. Clearly, any $n-1$ shares reveal no information about the secret. However, to achieve active security, this is not enough to guarantee correct opening.

Message Authentication Code (MAC) is a value that can be used to confirm a message has been created by a certain party who knows the MAC key, and to detect if a message has been changed. The main tool for achieving active security in modern, secret sharing-based protocols is information-theoretic MAC, as mentioned in the \cite{damgaard2012multiparty} \cite{damgaard2012multiparty}. A typical information-theoretic MAC scheme is listed as Alg.\ref{alg:MAC}.

\begin{algorithm}[ht]
\caption{MAC Check Procedure}
\label{alg:MAC}
\begin{algorithmic}[1]
\Require global MAC key shares $\alpha_i$, public sets of opened values ${a_1, \dots, a_t}$, associated MAC value shares $\gamma(a_j)_i$
\Ensure succeed if $\Call{MACCheck}{}$ passed, or $\varnothing$ if inconsistant MAC value is found.
\Method
\State All parties samples a seed $s_i$ and ask $\Call{Commit}{}$ to broadcast $\tau_{s_i}$.
\State All parties open $\tau_{s_i}$ and add up to get an agreed seed $s$.
\State All parties sample random vector $\textbf{r}$ on $s$
\State Each player computes the public value $a \gets \sum_{j=1}^t r_j \cdot a_j$. 
\State Player $i$ computes $\gamma(a)_i \gets \sum_{j=1}^t r_j \cdot \gamma(a_j)_i$, and $\sigma_i \gets \gamma_i - \alpha_i \cdot a$.
\State Player $i$ asks $\Call{Commit}{}$ to broadcast $\tau_{\sigma_i}$.
\State Every player open $\tau_{\sigma_i}$.
\State If $\sigma_1+\cdots+\sigma_n= 0$, the players output $\varnothing$ and abort.
\end{algorithmic}
\end{algorithm}

The security requirement is that the verification algorithm should succeed if and only if $m$ is a valid MAC on $x$, except with negligible probability. The information-theoretic property of the MAC scheme means that security holds even for an unbounded adversary. The MAC scheme we deployed in computation network derives purely from information theory. In other words, it cannot be broken even if the adversary had unlimited computing power. The adversary simply does not have enough information to break the encryption.

MAC plays a crucial role in making MPC verifiable. MAC check scheme makes sure that malicious party in the group can not bypass the security check by blending the verification value into the computation process.

\subsubsection{Revealing outcomes}

Once we have the MAC representations, it is fairly easy to describe the online phase of an MPC protocol in the preprocessing model. Since both of the authenticated secret-sharing methods described are linear, the parties can perform linear computations by simply computing locally on the shares. We can also add a public constant c to a shared value $[x]: P_1$ adds c to her share, and each party $P_i$ adjusts their MAC by $c\cdot \alpha_i$.

To multiply two secret-shared values, the parties need a random authenticated multiplication triple from the preprocessing phase. We also need some additional preprocessing data for sharing inputs, in the form of random authenticated masks where only the party providing input knows the value. When outputting a result of the computation the parties must check the MACs using the MAC check procedure described previously, which we denote by $\prod{MACCheck}$.

In summary, the computation process can be concluded as the protocol Alg.\ref{alg:computePhase}

\begin{algorithm}[htbp]
\caption{Computation Phase Protocol}
\label{alg:computePhase}
\begin{algorithmic}[0]

\Initialize The parties first invoke the preprocessing to get the shared secret key $\alpha$ , a sufficient number of multiplication triples $(\langle a \rangle,\langle b \rangle,\langle c \rangle)$, and random values $\langle r \rangle$, as well as single random values $\langle t \rangle$, $\langle e \rangle$. Then the below steps are performed according to the compiled function structure.
\Input To share secret input x of party $P_{input}$, involved parties take a random value $\langle r \rangle$. Then do the following:
\begin{enumerate}[1.]
    \item $\langle r \rangle$ is opened to $P_{input}$.
    \item $P_input$ broadcasts $\epsilon \gets x-r$.
    \item The parties compute $\langle x \rangle \gets \langle r \rangle + \epsilon$.
\end{enumerate}
\Add To add $\langle x \rangle, \langle y \rangle$, the parties locally compute $\langle x \rangle+ \langle y \rangle$
\Multiply To multiply $\langle x \rangle, \langle y \rangle $, the parties do the following
\begin{enumerate}[1.]
    \item Sacrifice
        \begin{enumerate}[a)]
            \item $\langle t \rangle $ is opened publicly, take two triples for check
            \item Evaluate $t \cdot \langle c \rangle - \langle h \rangle - \sigma \cdot \langle f \rangle - \rho \cdot \langle g \rangle - \sigma \cdot \rho$, where $\rho \coloneqq t\cdot c-h, \sigma \coloneqq b-g$
            \item Sacrifice one triple and authenticate the other
        \end{enumerate} 
    \item Multiplication
        \begin{enumerate}
            \item Take one authenticated triple
            \item Evaluate $\langle z \rangle = \langle c \rangle + \epsilon \cdot \langle b \rangle + \delta \cdot \langle a \rangle + \epsilon \cdot \delta$, where $\epsilon \coloneqq x-a, \delta \coloneqq y-b$
        \end{enumerate}
\end{enumerate}
\Output The parties unite to reveal output $y$ in the presence of $\langle y \rangle$. Correctness of computation should be checked first. Then do the following:
\begin{enumerate}[1.]
    \item Generate random linear combination coefficient $r_i=e^i$. Then compute $a=\sum_j{r_ja_j}$, where $a_j$ are opened intermediates.
    \item $P_i$ commits to $r_i=\sum_j{r_j\gamma(a_j)_i}, \langle y \rangle, \gamma(y)_i$.
    \item Open $\alpha$.
    \item Check $\alpha \cdot a=\sum_i \gamma_i$,$ \alpha \cdot y = \sum_i{\gamma(y)_i}$. If MAC check passed, output $y\coloneqq \sum_i \langle y\rangle $.
\end{enumerate} 
\end{algorithmic}
\end{algorithm}

\subsection{Preprocessing Process}\label{sec:preprocessing}
The preprocessing model began with Beaver's ``circuit randomization'' technique, which allows any arithmetic circuit to be securely computed by randomizing the inputs to each multiplication gate, using a multiplication triple, which is a triple of secret shared values $\langle a \rangle, \langle b\rangle, \langle c \rangle$ where $a$ and $b$ are uniformly random (in a finite field) and $c = a \cdot b$. Given this, two secret shared values $\langle x \rangle$, $\langle y \rangle$ can be multiplied by publicly reconstructing $d = x - a$ and $e = y - b$ and then computing

Since $a$ and $b$ are uniformly random, revealing $d$ and $e$ does not leak any information on the inputs x,y. The key advantage of this approach is that any MPC protocol based on linear secret sharing that operates on circuits in a gate-by-gate manner can now be easily recast in the preprocessing model: the preprocessing phase simply consists of creating many multiplication triples, and in the online phase the circuit is securely evaluated using these triples. Note that additions and linear operations can be evaluated locally because the secret sharing scheme is linear. After evaluating the circuit, each party broadcasts their secret shares of the outputs and reconstructs the result. The preprocessing stage is completely independent of both the inputs to the function being computed and the function itself (apart from an upper bound on its multiplicative size), and creating triples is typically the bottleneck of a protocol. 

\begin{algorithm}[htbp]
\caption{Preprocessing Phase Protocol}
\label{alg:preprocessPhase}
\begin{algorithmic}[0]
\Initialize generates global MAC key $\alpha$ and distributed key $\beta_i$
\begin{enumerate}[1.]
    \item The players agree on the public key pk
    \item Each $P_i$ generates $\alpha_i$, $\beta_i$, and define $\alpha \coloneqq \sum_i{\alpha_i}$
    \item Each $P_i$ computes $e_{\alpha_i}=\Call{Enc}{\alpha_i}$, $e_{\beta_i}=\Call{Enc}{\beta_i}$, and zero-knowledge proves on ciphers
    \item Compute $e_\alpha=e_{\alpha_1} \boxplus \cdots \boxplus e_{\alpha_n}$, and $\Call{DistDec}{}$ it to each party.
\end{enumerate}
\RandomValues generates random values for secret sharing
\begin{enumerate}[1.]
	\item Each $P_i$ generates $r_i$, and define $r\coloneqq r_i$.
	\item Each $P_i$ computes $e_{r_i}=\Call{Enc}{r_i}$, and zero-knowledge proves on ciphers.
	\item Compute $e_r=e_{r_i} \boxplus \cdots \boxplus e_{r_n }$. and $\Call{DistDec}{}$ it to each party.
\end{enumerate}
\Triples generates multiplicative triples for online computation
\begin{enumerate}[1.]
    \item Each $P_i$ generates $a_i$, $b_i$, and define $a\coloneqq \sum_i{a_i}$, $b\coloneqq\sum_i{b_i}$.
    \item Each $P_i$ computes $e_{a_i}=\Call{Enc}{a_i}$, $e_{b_i}=\Call{Enc}{b_i}$, and zero-knowledge proves on ciphers.
    \item Compute $e_a=e_{a_i}\boxplus\cdots\boxplus e_{a_n }$, $e_b=e_{b_i}\boxplus \cdots \boxplus e_{b_n}$. and $\Call{DistDec}{}$ it to each party.
    \item Compute $e_c=e_a\boxtimes e_b$, and $\Call{DistDec}{}$ it to each party.
\end{enumerate}
\end{algorithmic}
\end{algorithm}

Considering the malicious security model, we introduced two methods to solve the following issues: First, Zero-knowledge is applied to prove that certain values lie within a certain bound in case that the adversaries introduces noise to pollute honest parties' input. Second, distributed decryption allows corrupted party to add noise in triples, where we use sacrifice to double check triples validation.

\begin{figure}[ht]
\centering
\includegraphics[width=12cm]{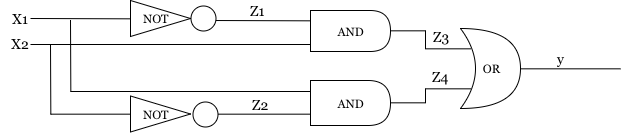}
\caption{Sample Arithmetic Circuit}
\label{fig:acircuit}
\end{figure}

Circuit used in this paper is the terminology describing evaluation functions in cryptography, but not the hardware circuit. To represent a computation, we can use either Boolean circuits or Arithmetic circuits. The main difference is with respect to their input's types and their gates types. It is nature to use Boolean notation when processing programs on bits and by Boolean operations, such as XOR, AND. Nevertheless, arithmetic circuits work on inputs in fields F by arithmetic operations, like field additions and multiplications. Figure \ref{fig:acircuit} shows a sample of arithmetic circuit.

\subsubsection{Compiling functions}

An MPC compiler translates a conventional function into an arithmetic or boolean circuit, in order for it to be accepted in MPC virtual machine. Corresponding to the normal compiler, which transforms high level language to assembly code, The MPC compiler turns the high-level function into MPC-friendly instructions.

The difference between arithmetic and Boolean function is not about functionality but the modality. The reason to choose one representative depends on the functions type to perform. Due to the mathematical basis, cryptography tools rely mainly on computation of field elements, which are treated as single input but not a string of bits. Then it makes sense to represent in arithmetic ways. However, many applications involve non-arithmetic operations like integer comparison, which is a basic procedure in many computations. In this case, implementing this non-arithmetic operation as an arithmetic one would be very inefficient, and Boolean circuits are a more natural representation.

\begin{table}[ht]
\centering
\begin{tabular}{*3c} \toprule
Circuits     &    Arithmetic      &     Boolean\\ \midrule
Elements     &    Bit \{$0$,$1$\} / $\mathbb{F}_2$  & Finite field $(\mathbb{F}_{p^k})^s$ \\
Underlying Operations   &    \Call{OR}{}, \Call{NOT}{}, \Call{AND}{}, \Call{XOR}{}   &   Addition, Multiplication  \\
Transform   &   Like an Arithmetic Logic Unit in processor   &  \begin{tabular}{c} 0 as addition neutral \\ 1 as multiplication neutral \\ $ \Call{OR}{a,b} = a+b-a\times b$
\end{tabular} \\ \bottomrule
\end{tabular}
\caption{Circuit Comparison}
\label{tab:circuit_comp}
\end{table}

The two representatives of circuits can be transformed to each other by an arbitrary translation shown in Table \ref{tab:circuit_comp}. The computable functions class of Boolean circuits of polynomial size correspond to the class of arithmetic circuits of polynomial size. Actually, we can picture arithmetic circuits as a generalization of Boolean circuits to arbitrary fields. There have been many papers that worked on extending cryptographic methods that work well for computations represented as Boolean circuit to the setting of computations which are best captured by arithmetic circuits, to get efficiency improvements for those circuits, or on bridging between methods adapted to both circuits, to capture computation whose best representation is a mix between the two.

\section{The Secure Computation processes} \label{sec:secure_computation}

\begin{figure}[ht]
\centering
\includegraphics[width=10cm]{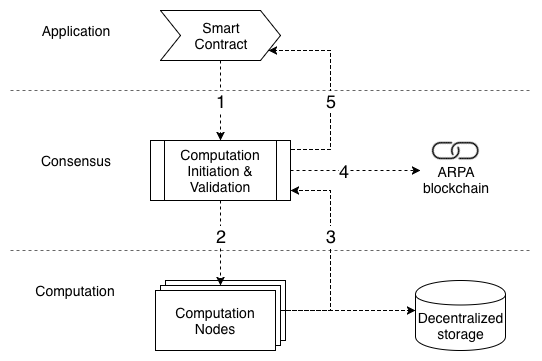}
\caption{The Secure computation process}
\label{fig:Secure_Computation_Process}
\end{figure}

The Secure Computation process starts from the application layer, typically by triggering a smart contract. To support off-chain computation, our network will observe smart contract events on a designated blockchain network, execute the computation work on ARPA, then submit results and proofs back to the smart contract. This process consists of 5 steps, as illustrated in Figure \ref{fig:Secure_Computation_Process}.

\begin{enumerate}
\item On the application layer, a smart contract representing a real-world contractual logic is initiated by a party who wants a heavy computation task or privacy-preserving task to be executed in the network while keeping the detail in private. The smart contract was created by the data provider (who charges a fee for providing information but doesn't want to expose the data to the public), or by several parties who want to engage in some private business logic.

To support off-chain computation, the smart contract could be one on another blockchain network. Our nodes can optionally observe the particular event defined in smart contract. And the result can be sent back to the smart contract by sending a message to the smart contract.

Once the smart contract received the deposit and other necessary information, such as data provider and function to be evaluated, the observing nodes will broadcast the event in the ARPA network, inviting other parties to join the MPC task.

\item All computation nodes who are interested in participating in the computation will start the bidding process. They are asked to deposit minimum required token to be eligible (detailed in Cryptoeconomics section). Next, a group of computation nodes are randomly selected, based on node's performance and condition, success rate, latency, etc. The random function used in this selection process is a decentralized random function, discussed in subsection \ref{nodes_selection}. 

One thing worth mentioning is that the data provider and data consumer are included in the group, making the computation nodes impossible to collude. This design is required so that at least one party is honest. Because our threshold protocol is secure up to n-1 parties, we can claim that this setup is \textit{cryptographically guaranteed to be secure}.

Once the quorum is determined, all bidding notes will know the result locally. This is a very important feature because even if the bribing attacker exists, he doesn't know whom to bribe or attack. The selected nodes don't need to broadcast the fact that they are selected, but instead they hold the cryptographic proof till the end of the process.

\item At the computation layer, the nodes in the pre-selected quorum coordinate together to perform ARPA MPC. As already described in previous section, they first secret-share the data. Then, a compiled function, in the form of boolean/arithmetic circuit is prepared for all parties. For each party, their job is simply to evaluate the function by following the circuit's instruction. During the computation, the parties actively exchange values on peer-to-peer communication, which is also secured with TLS. When the computation reaches the end, results and computation proofs will be produced. The computation proofs are Information Theoretic MACs produced alongside with the computation, i.e. they generated with the same computation steps as the result, making it impossible to forge. The results and MAC values will be signed with all nodes' secret keys to prevent forgery.

MACs are information-theoretic secure to convey the correctness of computation, so that the computation is verifiable. The validation complexity is $O(1)$, which means it consumes constant time to compute. Therefore it is possible to be executed in any VM. 

\item If the verification is passed, the payment will be processed, the gas fee will be distributed, and the meta-data of the transaction will be recorded on the blockchain. The ARPA consensus algorithm is discussed in section \ref{subsection:consensus}.

\item The result and MACs will also be sent back to smart contract. All nodes running smart contract will validate the authenticity of MACs, and then check the MAC's correctness. The detail of verification procedure is detailed section \ref{sec:secure_computation}. If the smart contract is running on another blockchain network, then other nodes will reach consensus by their own consensus algorithm.

\end{enumerate}

\subsection{Smart Contract}
A smart contract is executable code that runs on the blockchain to facilitate, execute and prove an agreement between untrusted parties without the involvement of a trusted third party. The benefit of a smart contract is that it can reliably transfer value, make decisions or perform computation as if there is a trusted third party. It also makes the whole executing process transparent, as anyone on the chain can audit the smart contract code and run the code to verify the results. The data processed or generated by a smart contract is also traceable in the receipt.

With many advantages, a smart contract has its limitations. The correctness of the smart contract is based on a network-wide consensus, which naturally requires every validating node to re-run it for verification purpose. This redundancy makes executing the smart contract very expensive, or in another words, the usability of the smart contract is limited. Another concern is privacy, as the smart contract itself and its data are transparent to every node on the blockchain network. 

Therefore, our approach to support off-chain computation for other chains is using smart contracts to carry the meta operations, including computation initiation, task broadcasting, and validation, so that the blockchain can reach consensus with computation proof. A pseudo-code example is provided to illustrate this concept. All of the heavy-lifting computation, including secret sharing and multiparty computation, are conducted by the MPC network.

\begin{lstlisting}[language=Solidity, caption=ARPA Smart Contract]
/** Written in Solidity-like pseudo-code. */
contract ArpaContract {
    struct Transaction {
        uint256 txnId;
        address dataConsumer;
        address dataProvider;
        string dataAddress;
        uint256 numParticipants;
        uint256 fund;
        bool isFinished;
    }
    
    struct MpcResult {
        // MPC may fail for various reasons, e.g. the fund is not enough to pay
        // compuation fee generated at run-time, or some parties aborted the
        // compuation.
        bool success;
        // MAC values are for MPC correctness verification.
        uint256[] macValues; 
    }
    
    mapping (uint256 => Transaction) public txns;
    uint256 public nextTxnId;

    event TransactionCreation(Transaction txn);
    event TransactionCompletion(Transaction txn);

    /**
     * Constructor function.
     */
    function ArpaContract() public {
        nextTxnId = 0;
    }
    
    /**
     * Data consumer calls this function to initiate an instance of MPC and
     * deposits fund for data usage and compuation fee.
     */
    function startTransaction(
        address dataConsumer,
        address dataProvider,
        string dataAddress,
        uint256 numParticipants,
        uint256 fund) public payable returns (bool success) {
        require(msg.value == fund);
        txnId = nextTxnId;
        nextTxnId += 1;
        txn = Transaction(
            txnId,
            dataConsumer,
            dataProvider,
            dataAddress,
            numParticipants,
            fund,
            /*isFinished=*/ false);
        txns[txnId] = txn;
        // Publish new transaction information to notify mpc network
        // to start compuation.
        emit TransactionCreation(txn);
        return true;
    }
    
    /**
     * The consensus layer of MPC network will call this function to report
     * completion of MPC and upload mpc result (including MAC values) for
     * verification.
     */
    function submitResult(uint256 txnId, MpcResult mpcResult)
        public returns (bool success) {
        txn = txns[txnId];
        require(!txn.isFinished);
        uint256 remainingFund = txn;
        
        if (verifyMpcResult(mpcResult)) {
            // Pay the fee to every party.
            remainingFund -= txn.dataUsageFee;
            txn.dataProvider.send(txn.dataUsageFee);
            for (worker : mpcResult.participants) {
                remainingFund -= worker.compensation;
                worker.address.send(worker.compensation)
            }
        }
        require(remainingFund >= 0);
        // If the data consumer deposited extra fund, or MPC verification
        // failed, refund the money to data consumer.
        if (remainingFund > 0) {
            dataConsumer.send(remainingFund);
        }
        
        txn.isFinished = true;
        // Publish new transaction information to notify mpc network
        // to start compuation.
        emit TransactionCompletion(txn);
        return true;
    }

    function verifyMpcResult(MpcResult mpcResult) public returns (bool success) {
        // Some MPC framework verifies MAC by checking (sum == 0).
        // This is only for illustration purpose.
        return mpcResult.success && sum(mpcResult.macValues) == 0;
    }
}
\end{lstlisting}

\subsection{Nodes Selection} \label{nodes_selection}
The nodes who will be participating in the computation are randomly selected based on a decentralized random function(DRF). One good candidate of such function is RANDAO, which is also a proposed in Ethereum PoS system\cite{randao}.


Combined with randomness from DRF, participating nodes are also selected based on the node's performance, condition, success rate, latency, etc.

\subsection{Off-Chain Computation}
Currently on-chain consensus is reached when everyone is receiving the same result, by executing smart contracts on each individual VM. There is no final authority when decentralized. Everyone has to validate every transaction in order to agree with others. The most prominent problem is the amount of gas consumed for every block. The reason that everyone has to repeat the same calculation is that blockchain is a trustless network and everyone must do the same and reach to the same result to trust others. In other words, there is no way to verify the result without actually going over the whole process, i.e. executing the code, in a trustless network.

Computation-intensive calculations are expensive on Ethereum and solidity is not good at expressing complex functionality. One way to compute complex function is to simply run it off-chain. Various solutions\cite{plasma, truebit} are focused on how to prove the correctness of an off-chain computation.

In order to bring computation off-chain, we need a verifiable scheme in the computation delegation process. For example, if one wants to prove that they have been somewhere, the best proof is an item that is unique to that place, such as a souvenir. We can think of off-chain computation as a delegation process, and the envoy has to prove that he actually did finish the task. Similar to this concept, an off-chain computation scheme needs a verification scheme that:
\begin{enumerate}
    \item Can prove that the computation is faithfully done,
    \item The amount of work done cannot be exaggerated, and more importantly,
    \item The privacy of the content is guaranteed and nothing is revealed to the participants.
\end{enumerate}

\begin{figure}[ht]
\centering
\includegraphics[width=8cm]{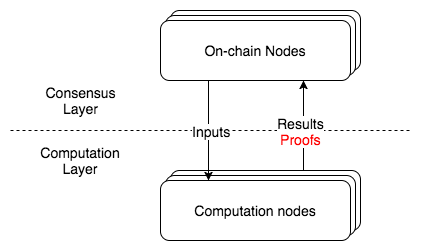}
\caption{Off-chain computation}
\label{fig:off-chain-computation}
\end{figure}

This relationship is illustrated in Figure \ref{fig:off-chain-computation}

We use Information Theoretic MAC to attain these desired properties. The following section will describe how MAC is used and why it can become the critical component in off-chain computation.

\subsection{Computation Verification}
Computation Verification scheme is the cornerstone of verifiable computation. As mentioned earlier, the current state of smart contract execution is duplicated across all nodes on blockchain. However, using MPC, the computation is guaranteed by mathematics and can be safely delegated to another set of nodes. In other words, the process can be trusted by a third party, as long as they provide proof that the process is done in a correct way. This question can be further broken down into three conceptual parts:
\begin{enumerate}
    \item The computation is conducted with prescribed function and correct input data,
    \item The proof cannot be forged, and
    \item The proof can be verified by other nodes.
\end{enumerate}
Under our MPC protocol, we are securing up to n-1 malicious notes using MAC. As described earlier, MAC is a value that can be used to confirm a message has been created by a certain party who knows the MAC key, and to detect if a message has been changed.

While the Smart Contract is being executed, a secure computation request is sent out from an external blockchain node to the computation layer. The computation network then selects several nodes for the task, and data provider's node is also included in the group. If any other party deviates from the protocol, such as sending incorrect result, the proof-carrying data will not be reconciled and the result will be discarded. Therefore, part 1 is guaranteed by our MPC protocol cryptographically.

After each party in the computation group finishes their work, they sign their MAC with their own private key, and store them on the DHT-based distributed storage. This prevents the MAC from being forged by a third party. Therefore, the proof cannot be forged and can be trusted to be authentic, even on a publicly-accessible storage. This produces desired property of part 2.

On the consensus layer, it is easier to claim that nodes should not trust other computation proof if they did not participate in the actual computation. Since the computation nodes was selected randomly, it takes all $n$ nodes to be colluded to pass the MAC check, which is a negligible probability or extremely high cost for the malicious party to control almost all the nodes in the network. And in such case, the network is probably does not exist anymore. In other words, the probability that can one trust the consensus network but not the computation network is not plausible.

Therefore, under the protocol, we know that if MAC does not pass the check, it means the computation was wrong because some node did it incorrectly. On the opposite, if the MAC does pass the check, we know that the computation was done correctly and the result can be trusted.

In summary, we now can conclude that the secure computation can be broken down into computation and verification parts. Further, we are able to prove that by using a privately-signed MAC, we can trust the related result, as well the identity of the signer.

\subsection{MAC with Off-chain computation}
In our off-chain computation phase, MAC is used in secret sharing schemes to ensure that corrupted parties cannot lie about their share and cause an incorrect value to be opened during reconstruction. The following equation shows how MAC proves correctness and integrity of computation. The protocol is under a full threshold active security model where security is guranteed even there is one honest party. This particular party can be data provider or computation initiator who has natural motivation to follow rules.

An authenticated secret value $x \in \mathbb{F}$ is defined as the following:
\begin{equation}
\langle x \rangle = (x_1, \dots, x_n, m_1, \dots, m_n, \alpha_1, \dots, \alpha_n)   
\end{equation}

where each party $P_i$ holds an additive sharing tuple $(x_i, m_i, \alpha_i)$ such that:
\begin{equation}
x=\sum_{i=1}^n{x_i},  x\cdot \alpha=\sum_{i=1}^n{m_i}, \alpha =\sum_{i=1}^n{\alpha_i}.
\end{equation}

When open a value $\langle x \rangle$, all players do the following protocol:
\begin{enumerate}
    \item Compute $\sigma_i=m_i-\alpha x_i$.
    \item Call \Call{FCommit}{} with (Commit, $\sigma_i$) to receive handle $\tau_i$
    \item Broadcast $\sigma_i$ to all parties by calling \Call{FCommit}{} with (Open, $\sigma_i$).
    \item If $\sum_i{\sigma_i}\neq 0$ then abort and output $\perp$; otherwise continue.
\end{enumerate}

The protocol achieves the integrity and correctness of computation by information-theoretic security of secret sharing and the following property.

\begin{equation}
\begin{aligned}
\sum_i{\sigma_i} &= \sum_i{(m_i - \alpha x_i)} =\sum_i{m_i} -\alpha\sum_i{x_i} \\
\sum_i{\sigma_i} &= \sum_i{(m_i - \alpha_i x)} = \sum_i{m_i} -x\sum_i{\alpha_i}
\end{aligned}
\end{equation}

It can be observed that, when given a MAC global key or value $x$, the protocol can check the result with MACs. These two ways can be used to ensure the correctness of computation by checking on intermediate and result by checking on output. Now we can see how MAC helps to protect the computation when at least 1 party act honestly.

Furthermore, we can also find that the broadcast value $i$ leaks no information on either secret value or MAC value because they are not correlated. Therefore, it is possible to transfer this check from computation layer to consensus layer, which helps to reach consensus on-chain. Also, the final summation of $i$ is mathematically simple to deploy any kind of virtual machine. In conclusion, MAC check is the ideal operation to promote consensus and protect computation privacy.

\textbf{Public auditable MPC} \cite{baum2014publicly} lift verifiability to arbitrary secure computations. Auditability means that even if \textit{all} the servers are corrupted, anyone with access to the transcript of the protocol can check that the output is indeed correct. This characteristic enables the separation of the input parties, computing parties, and the validator. After the protocol is executed, anyone acting as the validator can retrieve the transcript of the protocol from the bulletin board, and determine if the result is valid or not by using only the circuit and the output. The protocol will instruct computing parties to make part of their conversation public and logged on the bulletin. The verification is reached by linear commitments on secret inputs instead of traditional use of non-interactive zero knowledge proof.

The more ambitious properties that public auditable MPC is willing to achieve are universal verifiability, where the validators must not know the output and observe any process of the computation, but still hold computing parties accountable by Pedersen commitment.

\subsection{Data Storage}
\begin{figure}[ht]
\centering
\includegraphics[width=6cm]{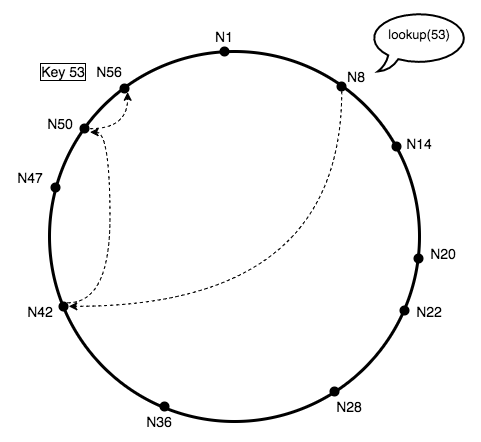}
\caption{DHT lookup}
\label{fig:DHT}
\end{figure}

The Distributed Hash Table (DHT) is a decentralized distributed storage system that provides a look-up service similar to a hash table. The data is distributed over the whole network of nodes, while any data is searchable by querying any node. To achieve that, the DHT (Figure \ref{fig:DHT}) defines a key space (for example, 160-bit string) and maps the entire data set onto this key space. The key space is then partitioned into chunks and assigned to nodes so that a node is only responsible for storing part of the data set. Another property is that every node keeps an index of mapping from key to the address of node(s) which stores the data with that key. With an overlay network connecting all nodes, a query that reaches any node can leverage the index to be routed to the node hosting queried data.

The decentralization and fault tolerance nature enable DHT to be more resilient to a hostile attacker than a typical centralized storage system. DHT can also be designed to have Byzantine fault tolerance\cite{lamport1982byzantine}. This makes it feasible to handle MAC value storage in ARPA's network, as no individual MPC node can be trusted. On the other hand, if we choose an arbitrary MPC node to store the MAC value, there is a chance that the node will supply a forged value, either because it happens to be a malicious node or it is under hostile attack.

\section{Cryptoeconomics} \label{sec:crypto}
Bitcoin is considered the first massive deployment of a Byzantine Fault Tolerant (BFT) protocol in real world. Byzantine failures are considered the most general and most difficult class of failures among the failure modes. In 1999, the Practical Byzantine Fault Tolerant(PBFT)\cite{pbft} was introduced and received immediate notice. The algorithm is designed to work in asynchronous systems and is optimized to be high-performance, with an impressive overhead runtime, and only a slight increase in latency. However, the solution works on a permissioned system, meaning that the nodes have to be authorized to join the network. This property works well on closed blockchain system like Hyperledger\cite{hyperledger}, Steller\cite{mazieres2015stellar} and Ripple\cite{schwartz2014ripple}.

The Bitcoin network, on the other hand, is a permissionless network. It works in parallel to generate a blockchain allowing the system to overcome Byzantine failures and reach a coherent global view of the system's state. To achieve decentralized consensus on a permissionless network, Bitcoin invented a new consensus scheme called proof-of-work. This consensus scheme has three major properties:
\begin{enumerate}
   \item The "work" is a NP hard problem to prover but easy to verify by verifier.
   \item The more stake (computation device) and cost (electricity) a node is willing to put into the system, the higher the chance that this node will be the first to solve the work.
   \item The node who solves the work first will be rewarded and its ledger is broadcasted and trusted.
\end{enumerate}

The prover, who found the hash value that satisfies the nonce, can be trusted because he puts tremendous efforts/cost/stake in this game, and it would not be a rational decision to hack the system. Otherwise he gained nothing with depreciating token value.

The difference between PBFT and Bitcoin is that the former makes an honest assumption of participants (a permissioned network), while the later gives a more realistic and real-world assumption of \textit{Economic Rationality}\cite{rationality}, a concept from neoclassical economics, which is maximization of subjective utility. Chain-based blockchain systems like Bitcoin or Ethereum prefer plausible liveness to correctness, while BFT consensus algorithm can ensure transactional correctness but may not be able to finalize conflicting blocks and can get stuck. 

In Ethereum and other blockchain networks based on the Nakamoto(Proof-of-Work, or PoW) consensus, the fundamental assumption is that the computation resource is scarce. If everyone is verifying transactions, there will be no final state of transaction history agreed by everyone. If a leader in the network propose a block of transaction and everyone else only has to verify that, the situation is alleviated. Therefore, consensus is reached based on block of transactions, proposed by a participating node who solved a cryptographic puzzle using fairly large amount of computation power. This brings a trustworthy "finality" to the blockchain network. In essence, Bitcoin created a game where verification (of the hash work) is much more simpler than the computation of the function(verify the transactions). Nakamoto consensus protocol transfers the trust needed to verify transactions to a puzzle game. With \textit{economic rationality} mentioned ahead, the finality of the state is set by a trusted leader who solves the hash puzzle. 

Similar to how Bitcoin reached decentralized consensus, our work makes \textit{Economic Rationality} assumption to MPC, so that the cryptographic protocol can be deployed as a permissionless network. On the contrary to the assumption of Proof-of-Work consensus, ARPA designed the MPC protocol where verification of computation is even simpler than hash verification, using information theoretic MAC. This feature opens new possibilities of consensus, explained in the \textit{Consensus}(\ref{subsection:consensus}) section.

\subsection{Economic Rationality}
We aim to deploy an MPC network with cryptoeconomics considered. While in cryptography the most basic assumption is that some parties are simply honest (or not), in economics the assumption is that the parties are rational and their utilities can be quantified, such that with a proper incentive mechanism all (or at least most) of them can be discouraged from deviating the protocol.

Under rational behavior, everyone will attempt to break from the rule and gain as much as possible. But to a point where no more economic benefits one can gain. If the situation occurs where deviating the protocol, or even sabotaging the whole network, is in the best interest for the participating nodes, they would do so for their own benefit. On the contrary, participants will follow the protocol if the incentive received when following protocol is more than expected return from breaking it. 

We designed a incentive system that will punish those who deviates from the protocol, and the punishment is more than the expected return when colluding with others. This will eliminate participating nodes to take chance in the system. This system will be explained in detail in section \ref{accountable}.

Sybil attack is common in peer-to-peer network, in which spawning identities will help the attacker to gain disproportional advantage on voting or influence. In PoW consensus scheme, the spawned nodes are bounded by the real computation power on their physical machine. Therefore, the risk of sybil attack is mitigated. 

We designed a credit system that each node is in multiround game system, and it is the best interest for them to keep the same identity (than register a new address) and follow the protocol. 

Without rationality, one can simply believe that the super admin of AWS (or anyone with unlimited computation resources) can attack Bitcoin and destroy its belief system. But in reality, it doesn't have to be worried because
\begin{enumerate}
    \item A rational person would not do so without a budget. 
    \item When invested with computation resources, a rational participant would rather make more economic return by staking in the game than sabotage a network with some upfront cost.
    \item A rational person will not assume that others would so that
\end{enumerate}
Therefore, it can be concluded that this sabotage behavior will never happen.

With a properly designed incentive system, one brings the question from a mathematical (absolute) perspective to a more realistic angle, where every participant can be "guided" to follow the protocol, and those who don't will be punished.

\subsection{Identity and Reputation}
At the time a new node joins the computation network, it is allocated a unique account in the consensus layer. The account will be served as public identity of such node. With asymmetric encryption, the network can generate private/public key pairs as the identity for a node. The public key serves as the node identification and information authentication. To prove the authenticity of the result, one has to sign the MAC value with its private key.

\subsection{Accountable safety}\label{accountable}
The computation node participating in the computation has to put up a stake in order to participate in a computation task. 

We require the staking value to be: $$x = max(computation\:stake, intel\:value\:stake) \cdot security \:multiplier $$
where:
 $$computation\:stake=(total\:computation\:resources\:estimated) * (n-1) $$
and $Security\: multiplier$ can be a constant determined by the network

There are several ways to break the protocol and be a malicious node. Each case is listed and we will analyze how our system can handle such behavior.

\paragraph{Not following the circuit}: functions are compiled in an arithmetic circuit for the nodes to compute jointly. If a node does not follow the instruction, the result will be incorrect. However, since we require MAC as a proof of computation, and the proof is tightly linked to the computation process, we can immediately tell if the result deviates from the assigned function by verifying the MAC all together. Using MPC validation, there is actually no point of doing so. Another argument of not doing so is that, since the computation is heavy, it makes more sense to just abort the task if one wants to sabotage the computation, in which case we will refer to case 4).

\paragraph{Sending out incorrect result:} as detailed in MAC session, it is impossible to send the wrong value with the correct MAC, as the MAC is paired with the result and can be detected.

\paragraph{Sending out incorrect MAC:} the MAC cannot pass the validation process and there is no benefit of doing so.

\paragraph{Abort attack:} similar to DoS (Denial of Service) attack, the node simply stops in the middle of the collaborative computation. Under such case, we can set up a timeout parameter for the computation task. If one party in the computation task fails to return before the timeout, then the task can be deemed as failure. The party then should compensate others the cost of computation resources already spent in the task. A stake larger than $(total\: computation\: resources\: estimated) \cdot (n-1)$ should be required at the election phase to prevent abort attack.

\paragraph{Stealing the data:} MPC is designed to deal with this problem – how to compute a function with distrusted parties without leaking the data. Our MPC protocol is designed to be secure up to $n-1$ colluded parties. In our case, in addition to the original $n$ party who agreed to compute, we will also select $m$ additional random nodes to increase the security. One has to collude all $n + m$ parties to steal information. This is impossible because the original data owner cannot be colluded. 

In sum, our staking mechanism is designed in such a way that, it is in the best interest of each participant to follow the protocol. This is due to the greatest benefit of the malicious action being lesser than the greatest expected loss of following the protocol. 

\subsection{Uncoordinated majority}
We assume that all nodes are rational in a game-theoretic sense, but no more than some fraction (often between 25\% and 50\%) are capable of coordinating their actions. The Proof-of-Work in Bitcoin with Eyal and Sirer's selfish mining fix is robust up to 50\% under the honest majority assumption, and up to 23.21\% under the uncoordinated majority assumption\cite{eyal2018majority}.

Under ARPA's MPC framework, with $t-1$ unconditional security, coordinated majority is unrealistic, because the . Because the colluded party needs to collude almost all nodes to be able to control the network. See Table \ref{tab:coordinated_chance}, where a coordinated attack requires all parties in a MPC computation to be colluded. With even 95\% of the nodes are colluding together, they still have less than 1\% chance to be able to break the security protocol.

\begin{table}[ht]
    \centering
    \begin{tabular}{*3c}   \toprule
      Coordinated\%   &  Quorun   &    Chance \\   \midrule
       50\%            &  100      &     0.00\%  \\
       60\%            &  100      &     0.00\%  \\
       70\%            &  100      &     0.00\%  \\
       80\%            &  100      &     0.00\%  \\
       90\%            &  100      &     0.00\%  \\
       95\%            &  100      &     0.59\%  \\
       98\%            &  100      &     13.26\%  \\
       99\%            &  100      &     36.60\%  \\  \bottomrule
       
    \end{tabular}
    \caption{Coordinated chance}
    \label{tab:coordinated_chance}
\end{table}

Clearly, it is unrealistic to make economical benefit when 95\% nodes are from the same party. Because it is virtually hosting a private network and hope to steal token issued on its own.

In addition, in most cases, the data provider will be a part of the MPC group. And because the data provider will not collude with others, there is zero chance that the colluding parties will break the protocol and get the secrets.

Similar to PoW, joining a MPC network requires a decent computer in order to be able to earn reward on MPC tasks. This investment prevents attacker to flood the network with virtual nodes.

\subsection{Bribing attacker model}
Attacker is motivated to make the network fail in some way. Attacker can be one of the participants, or the attacker can sits outside the protocol, and has the ability to bribe any participants to change their behavior. Attackers are modeled as having a $budget$, which is the maximum that they are willing to pay, and we can talk about their $cost$, the amount that they end up paying to disrupt the protocol equilibrium.

As discussed at section \ref{nodes_selection}, the participating nodes are selected at random, and the selected party will not be publicly known until they have completed the computation. Therefore, there is no target for the  attacker to bribe with.

Without a bribing target, an attacker needs to bribe more than 95\% nodes to make a substantial impact on the network. Even the attacker can wait passively until the chance to appear, the cost to bribe the whole network is much more than a reasonable attacking budget and thus rendered unrealistic. 

With a proper penalty designed, a hidden bribing target, and a high threshold to bribe the whole network, ARPA's network can still be secure under this security model.

\subsection{Consensus Algorithm} \label{subsection:consensus}
The blockchain consensus algorithm mainly solves two problems - who is responsible for mining the block and what to do when there is a fork. Difficulties usually lie on how to solve the fork, that is, when there are two or more legitimate forks, how to let everyone agree on one chain and continue, which requires a unified standard of judgment. Nakamoto consensus chooses the most expensive fork with the highest cost, and verification of the most expensive chain has to be objective and verifiable, which is, the Proof-of-Work. It does not mean that there is only one way to solve the fork. Such as God's dice, where everyone agrees on the fairness of the dice. Whenever there is a fork, the dice can decide on a fork. This consensus can also ensure the normal operation of the blockchain system.

How to design a practical and verifiable fair random number generator based on the blockchain has become an important research problem in recent years. There were programs \textit{Randao} proposed by Vitalik Buterin\cite{randao2016tor}, some DApps that used Oraclize to obtain the random numbers from the off-chain service, to implement a blockchain random number generator. The Ethereum Foundation also listed random number generation on the blockchain as an important issue to be researched for the next 2 to 3 years\cite{ethereum2018randnum} and invited all parties to participate on a solution.

In ARPA's network, consensus is designed with a two step development plan. Phase 1 is a PoW consensus scheme, where a new block is found and broadcast by a mining node who find a hash result satisfying the nonce of the network. A phase 2 consensus is reached by a lottery system, where participants purchase tickets a few blocks ahead to enter the candidate pool. At block creation time, the candidates together calculate a pseudo-random function, using block header and other transitional information as the seed for the random function.

\subsubsection{Nakamoto Consensus}
ARPA is dedicated to make a general-purpose secure computation network available for the first time in history. While Proof-of-Work consensus is known for its waste of energy, its stableness and secureness are still being recognized and used in most major blockchain networks. 

At the beginning of our development phase, ARPA's team will focus on solving the cryptographic challenges in MPC protocols and computation layers of the network. Rather than being innovative at every corner of our first release, a decent and proven version consensus algorithm will be used at our first release of mainnet. Therefore, a Proof-of-Work consensus algorithm will be chosen so that we can spend more efforts to the development of secure computation layer and make it really robust. 

We will use multiple hashing functions at random to prevent ASIC chips mining in our network, because we prefer participating nodes joining ARPA to have  decent computing power using CPUs (the capability desired by MPC). Choosing hashing function randomly at each block guarantees that the mining nodes would be able to conduct general-purpose computation with the capability of CPUs, one that cannot be reached by ASIC chips.

\subsubsection{Lottery Consensus}

The pseudo-random function must have the following characteristics: deterministic, provability, uncontrollability, unpredictability and Tamper-resistant to attacks.
\begin{itemize}
    \item Deterministic: given all information revealed, the result will be determined and unique.
    \item Provability: The result can be validated by anyone after the information was revealed.
    \item Uncontrollability: The result cannot be influenced by anyone, or a partially colluded group.
    \item Unpredictability: Using transitional information such as block header, time, participants address, will result unpredictable random result.
    \item Tamper-resistant: The result cannot neither be calculated ahead of time, nor be forged with huge amount of computation power.
    \item Privacy: The random result is opened but only the winner knows he won the lottery. Therefore the attacker doesn't know the target until the winner broadcasts the signed validation.
\end{itemize}

Several choices for random generator are: RANDAO, BLS-based multisig, or Secret-sharing based pseudorandom. All can achieve the desired characteristics mentioned above. Our protocol is designed as follows: At block time $t$, the nodes who want to become the validator, purchase a ticket to be eligible and submit a hash proof. At time $t+x$, where $x$ is the waiting period, each party generates $p_i$ corresponded to the hash committed before. We formally describe the procedure in Alg.\ref{alg:DRF}

{\centering
\begin{minipage}{.7\linewidth}
\begin{algorithm}[H]
\caption{Decentralized Random Function}
\label{alg:DRF}
\begin{algorithmic}[1]
\Method
\State Decide $n$ participants in the game
\State $p_i \gets \Call{sha3}{s_i}$: Output a hash value by random value $s_i$, related to transnational information.
\State $P \gets \sum p_i$: Open value $P$ and reveal secret $s_i$
\State $0/1 \gets \Call{Ver}{s_1, s_2 \dots s_i}$: Check if the $Commit$ on $s_i$ is correct. The result of $P$ is valid if honest check is passed for everyone.
\end{algorithmic}
\end{algorithm}
\end{minipage}
\par
}

 If the authentication check passed, we calculate $r = R\: MOD \:n$, and obtain the randomly selected group of nodes. One of the nodes will become the $prover$, and the rest will become the $validator$. The $prover$ will create a block containing all recent valid transactions, and the $validator$ will challenge the $prover$ by checking the validity of the new block. If $prover$ failed to create a valid block, it will lose its deposit. Otherwise, the $prover$ will be rewarded with a small fee, along with the ticket price.

\subsection{Token Usage}
ARPA tokens, the native token for ARPA, perform five interconnected and critical functions within the ecosystem, over dispersed, dynamic networks:

Medium of Exchange – all transactions within ARPA computation network use ARPA tokens as the medium of exchange for services. For example, data providers receive token payment from data consumers, while model providers receive token payment from data owners

Computing Cost – ARPA tokens are exchanged to perform secure multiparty computation and compensate for computation providers. Computation cost is measured in triples used during the process and other factors

Stake – computational power providers will use the ARPA token as a form of safety deposit for launching and fulfilling computation jobs. Abort during computation and other malicious actions result in loss of stake. This will ensure fairness for all parties and limit misuse of ARPA by bad actors

Transaction Fee – similar to gas fee on Ethereum, ARPA tokens are paid to nodes facilitating the transaction as a fee

Backing of Data or Model – all audiences of ARPA's network, ranging from normal crypto audiences and network stakeholders to professional investment entities, can back public data or model following Additive Backing System. Additive Backing System broadens ARPA's audiences and incentivizes early adopters and backers of ARPA's data marketplace.

\section{Experiments}  \label{sec:experiments}
Various experiments are conducted in privacy-sensitive applications, like auction or machine learning. The performance of the practical results shows that under suitable optimization and diligent function compilation, the overhead caused by computation and communication can be largely mitigated. With such effort, the security and privacy brought by MPC is worth the price of only 2-3 magnitudes of additional computation power.

\subsection{Vickrey Auction}
As for the high party number case, a secure Vickrey second price auction is implemented on AWS instances, where 100 parties input one bid each. The auction system will reveal the winning bidder and the second-high bid. The Vickrey auction requires 44,571 triples. The performance is listed in Table \ref{tab:vickrey}.

\begin{table}[ht]
\centering
\begin{tabular}{ *3c } \toprule
 Party number & Offline phase & Online phase \\ \midrule	
 100 & 98 s & 1.4 s \\  \bottomrule
\end{tabular}
\caption{Vickrey Auction Performance}
\label{tab:vickrey}
\end{table}

\begin{figure}[ht]
\centering
\includegraphics[width=10cm]{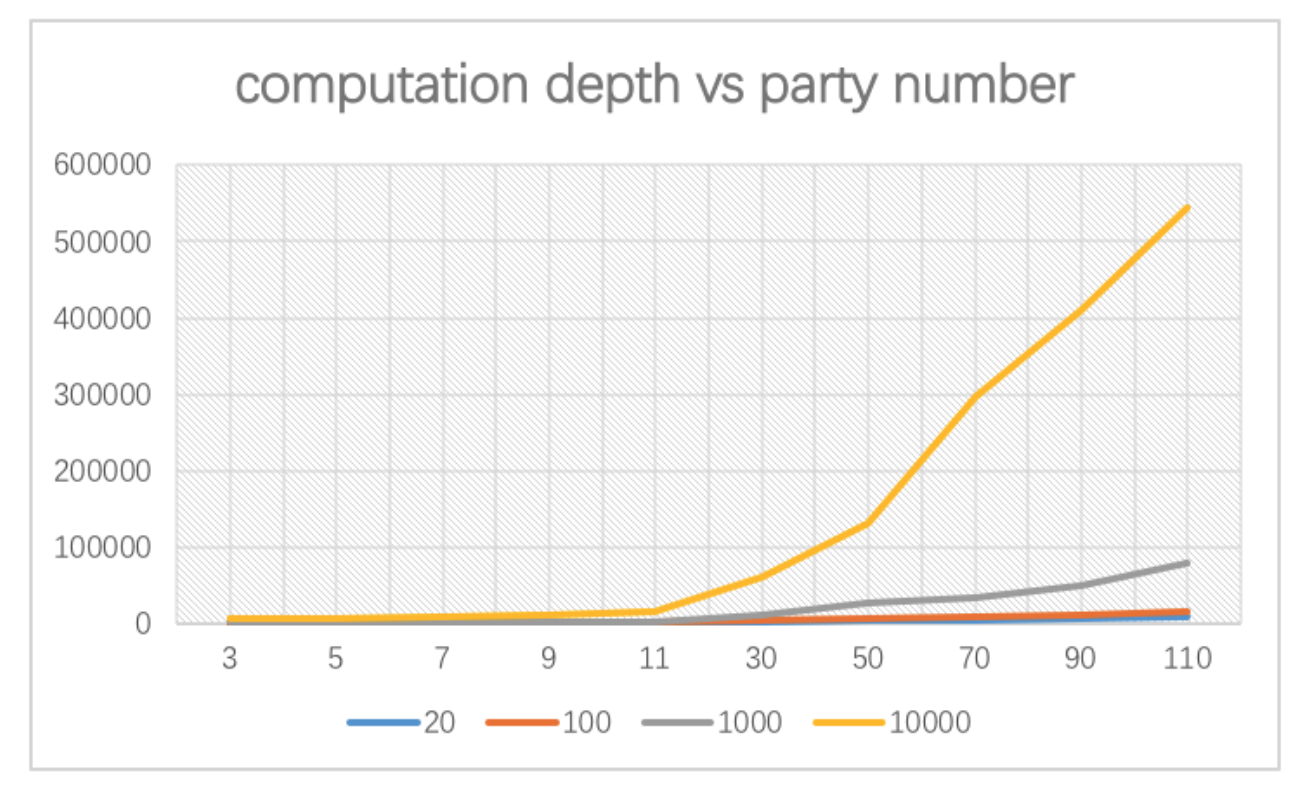}
\caption{Computation overhead in Vickrey Auction}
\label{fig:computation_overhead}
\end{figure}

To make it more general. An evaluation experiment on different circuit depth and party numbers is performed. The underlying field is set as 61-bit Mersenne field, security is approximately 260. The test is done for different numbers of parties on a series of circuits of different depths, each with 1,000,000 multiplication gates, 1,000 inputs wires, 50 output wires. The circuits had 4 different depths: 20, 100, 1,000 and 10,000. The experiment was run on AWS in a local network configuration in North Virginia region AWS. Each party was run in an independent AWS C4.large instance (2-core Intel Xeon E5-2666 v3 with 2.9 GHz clock speed and 3.75 GB RAM). Each execution (configuration, number of parties, circuit) was run 5 times, and the result reported is the average run-time. The computation overhead is depicted in Figure \ref{fig:computation_overhead}

\subsection{Deep Learning}
We will describe the implementation of Convolutional Neural Networks (CNNs) using MPC. The biggest problem here, when training large neural networks, might be the maturity and availability of arbitrary precision arithmetic on GPUs. Although some hope exists, this area is still under developed and needs further exploration. In this experiment we will still use CPU as an execution environment.
Our MPC approach uses secret-sharing to mask the values used in the computations. This allows for arithmetic circuit evaluation, which is greatly more efficient for training neural nets. Both the ReLU and max pooling are nonlinear operations, which are expensive to compute in MPC. Therefore, the ReLU is approximated using a polynomial and the max pooling is replaced by average pooling. 
A simple implementation of a convolutional layer can be found here, which is used as a basis for our implementation.

\begin{lstlisting}[language=python, caption=Convolutional Neural Networks (CNNs) with MPC]
def conv2d(x, filters, strides, padding):
    # shapes, assuming NCHW
    h_filter, w_filter, d_filters, n_filters = filters.shape
    n_x, d_x, h_x, w_x = x.shape
    h_out = int((h_x - h_filter + 2 * padding) / strides + 1)
    w_out = int((w_x - w_filter + 2 * padding) / strides + 1)

    X_col = x.im2col(h_filter, w_filter, padding, strides)
    W_col = filters.transpose(3, 2, 0, 1).reshape(n_filters, -1)
    out =
      W_col.dot(X_col).reshape(n_filters, h_out, w_out, n_x).transpose(3, 0, 1, 2)
    return out
\end{lstlisting}

For the dot products, specialized triples are created, ensuring that every value used in a dot product is only masked once. And for the convolution for which holds: im2col$(A)\cdot B=C$. Therefore, only a masked version of both the inputs (masked with A) and the weights (masked with B) are communicated, without any duplicates. 

\paragraph{Reusing triples} When a variable A is used in two operations with B and C to create D and E, we can create triples such that: $operation_1(A,B)=D$  and  $operation_2(A,C)=E$. In the backward phase the error is typically used in two operations: once to compute the backpropagation error, for the previous layer, and once to compute the weight updates. Moreover, both the weights and the layer inputs have been used already in operations in the forward phase to compute the layer output, and can therefore be reused in the backward phase.
To gain insight, we compare the number of values communicated per participant in one iteration for a batch size of 128 images from MNIST of a simple two layer convnet with the optimized version. For the baseline, the backward phase is more expensive than the forward phase. For the optimized model, it is the other way around. This is mainly because we can reuse the masks created in the forward phase.

\begin{table}[ht]
\centering
\begin{adjustbox}{max width=\textwidth}
\begin{tabular}{ *9c } \toprule
 Layer & Forward & Backward & \multicolumn{2}{c}{Total} &Forward & Backward & \multicolumn{2}{c}{Total} \\ \midrule	
 conv2D(32,3,3) & 903K & 4,114K & 5,018K & (17\%) & 101K & 803K & 903K & (18\%) \\
 avg\_pooling2D(2,2) & - & - & - & - & - & - & - & - \\
 ReLU(approx) & 3,211K & 3,211K & 6,423K & (22\%) & 1,606K & 803K & 2,408K & (47\%) \\
 conv2D(32,3,3) & 7,235K & 8,840K & 16,075K & (54\%) & 812K & 201K & 1,013K & (20\%) \\
 avg\_pooling2D(2,2) & - & - & - & - & - & - & - & - \\
 ReLU(approx) & 803K & 803K & 1,606K & (5\%) & 401K & 201K & 602K & (12\%) \\
 dense(10,1568) & 216K & 219K & 435K & (1\%) & 216K & 1K & 218K & (4\%) \\ \midrule
 total & 12,368K & 17,188K & 29,556K & (100\%) & 3,136K & 2,008K & 5,144K & (100\%) \\
 
 \bottomrule
\end{tabular}
\end{adjustbox}
\caption{Communicated number of 128-bit values per iteration in the double conv layer architecture}
\label{fig:convnet}
\end{table}

\section{Ecosystem \& Applications} \label{sec:ecosystem}
In this section we discuss how business logic and applications can be built when secure computation is available. This can happen either on other blockchain system with the ARPA system enabled, or directly on ARPA's own blockchain.

\subsection{Design}

Developers can leverage ARPA to protect dApp data privacy and build privacy-preserving dApps, free of cryptography knowledge.

ARPA's secure computation also allows data to be shared without disclosing the raw data to anyone during data-at-use. ARPA can potentially disrupt the whole data industry by eliminating trust issues and data intermediaries within: finance, healthcare, IoT, retail, energy sectors, and other sectors where data is a valuable and sensitive asset to enterprises and individuals.

A reputation system is established on our data marketplace scenario, where data providers are able to sell data value without disclosing the raw data, and a data consumer can discover and use the data without signing an NDA contract with the data provider.

A node's reputation increases by being a reliable and responsible computation provider, and this reputation is permanently logged on the blockchain. After each transaction, each party can leave a review about data quality and model training results. Good reputation attracts more buyers and business, increasing the number of backers (we will discuss the incentive system later).

\subsection{Participants}
\subsubsection{Computation Nodes}
Computation nodes supply computing power and storage for multiparty computation. Computation nodes have to put up a safety deposit of ARPA's token(s) for computation tasks based on projected triple consumption. In case of abort, the node's safety deposit is compensated to other nodes in the same MPC task, as well as, to the next group of nodes performing the same task. 

Computation nodes have the following attributes:
\begin{itemize}
    \item Strong computing power to ensure fast triples generation during the data pre-processing phase of multiparty computation
    \item High availability and uptime during online computation, guaranteed by node's stakes
    \item Fast and reliable internet connection, governed by the node's stakes
\end{itemize}

\subsubsection{Data Providers}
Data providers can be either individuals, enterprises, research institutions, financial,  and medical institutions that possess valuable but under-utilized data. ARPA enables data providers to securely monetize their data with privacy-preserved computation. Data providers earn token by sharing out data value but not data ownership. ARPA's privacy-preserving feature for the data marketplace can potentially unlock massive amounts of data from enterprises, as well as, from individuals as a new way of data monetization. 
\subsubsection{Data Consumer}
Data consumers can be enterprises, researchers, government, universities, and individual developers that rent data to run models, monetizing on analysis output and training AI/ML models. Data consumers pay token(s) for access to data, and benefit from lower costs by renting data. Not by buying data from suitable enterprises or individuals. Data consumers can also leave ratings and reviews for a data-set after the transaction takes place. Essentially ARPA's privacy-preserving computation cuts out data aggregators and intermediaries that used to take majority of profit within data transactions. 
\subsubsection{Model providers}
Model providers can be enterprises and individual developers that own models and would like to monetize models based on usage. Model providers earn tokens renting out models and also get incentivized sharing models to ARPA's network. Models can be rented on a pay-per-use basis or by subscription. Without potential data leakage risks, data owners can screen for the best model available on ARPA, run the model with data in a cryptographically secure way, and get insightful output from the model provider. One use case can be running analysis for medical data from hospitals where a third-party AI developer directly processes encrypted medical data and sends back results to hospitals without access to raw data. 
\subsubsection{Data / Model Backers}
Data and model backers can be individuals and enterprises that invest in public datasets or models, and receive revenue shares in tokens. Data and model backers can range from normal crypto audiences and network stakeholders, to professional investment entities. Any dataset or model made available to the public at ARPA will mock up 20\% of the dataset selling price. This price difference will be distributed to data backers. It is generally hard to price datasets even when they are visible rather than encrypted, and therefore we introduce an Additive Backing System (ABS) to reflect data quality. ABS is designed to broaden ARPA audiences and attracts more early adopters of the network. 

The basic rules of ABS are:

\begin{algorithm}
\caption{Additive Backing System}
\begin{algorithmic}[1]
\Input $R_{share}$, $N_{backer}$;
\Output $R_{share}$, $P_{invest}$;
\Define Variable definition as follows;
    \begin{enumerate}
        \item $P_{invest}\coloneqq$ Investment amount of newcomer;
        \item $R_{share} \coloneqq$ Revenue share of each backer;
        \item $R_{backer} \coloneqq$ The revenue of all backers;
        \item $N_{backer} \coloneqq$ Index of backers;
    \end{enumerate}
\Method
\State $ R_{share}=R_{backer} / N_{backer} $ , $P_{invest}=N_{backer}$;
\State The $n$th person backing the dataset pays $n$ ARPA token;
\State The revenue share of dataset is evenly distributed to $n$ backers;
\State The $i$th person who can sell his position at $n+1$ ARPA token, given there are $n$ backers already.

\end{algorithmic}
\end{algorithm}

Based on the algorithm of ABS, we can now prove that the number of backers is positively correlated to the expected income of a dataset. A longer backing line represents better quality data, and it is a very visible indicator for a data consumer to find high quality datasets. Thus, we effectively outsourced the task of quantifying data quality to the ecosystem.

\begin{figure}[ht]
\centering
\includegraphics[width=10cm]{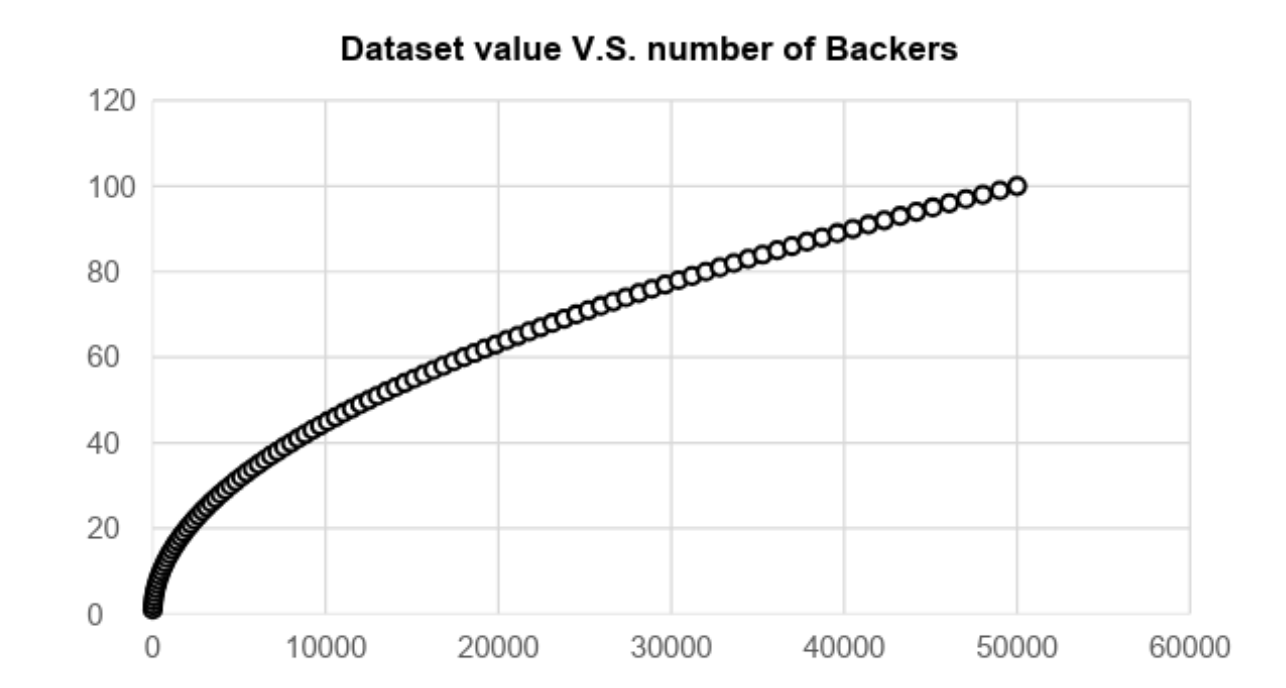}
\caption{ABS system}
\label{fig:abs}
\end{figure}

\subsection{Business Flow for Data Renting}

\begin{itemize}
    \item The data provider uploads a dataset and chooses to make it public or private (for a designated data consumer only). The Data provider sets the price on a per-use basis.
    \item Data backers can view data samples and invest in public datasets to collectively share 20\% of its future revenue. Backers are also able to see past transactions and reviews regarding the dataset. The amount of contribution is based on a Linear Backing Mechanism.
    \item A data consumer can request to run its model with the data provider's dataset. In addition to the dataset usage cost, there will be estimated computation cost required. 
    \item The smart contract ran by the data consumer, and a deposit will be required to initiate the MPC computation. Once deposit sends to the smart contract, the MPC task will be broadcasted.
    \item ARPA finishes the computation task. The data consumer and computation nodes get paid, and the data consumer gets the output data. Dataset backers also get their shares of the reward.
\end{itemize}

\begin{figure}[ht]
\centering
\includegraphics[width=13cm]{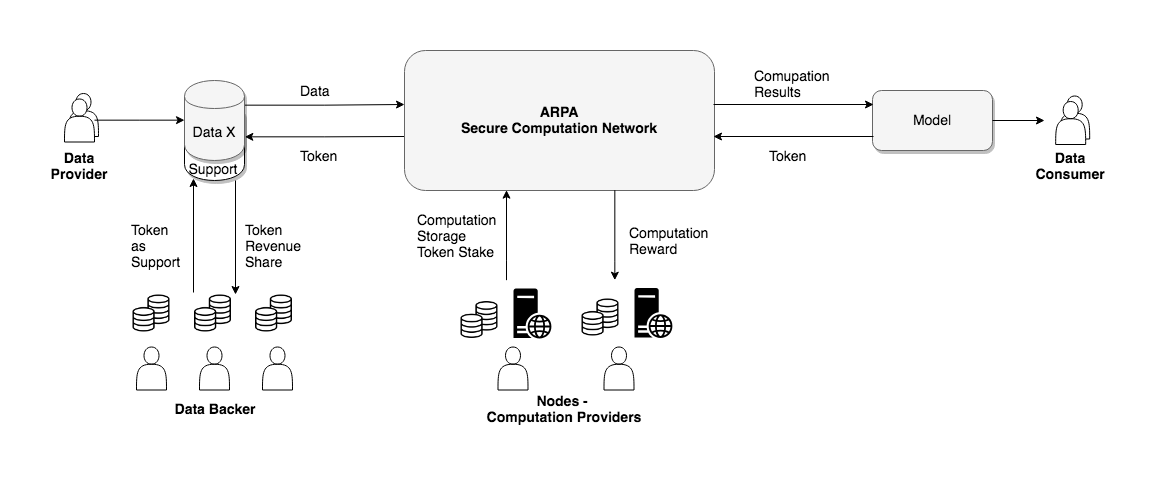}
\caption{Privacy-Preserving Data Marketplace}
\label{fig:data_market}
\end{figure}


\paragraph{Case 1: Sharing of Behavioral Data in Various Sectors}
A recent use case is data sharing and modeling as a service offered by banks to their corporate clients. For example, bank X is offering retail company Y data services by sharing the credit card transactions of all industries, as well as predictions on both individual and industry levels (e.g. customers' lifestyle, shopping preference and industry outlook). This data is expected to be merged back to company Y's customer data to be further used for various analytical functions. Currently the bank hashes PI data and then sends it to a 3rd party company to perform data matching with retailers' data. 

ARPA's secure multiparty computation network can effectively eliminate 3rd party data matching intermediaries by secret-sharing a bank's and retailer's data to computation nodes, performing data matching with bank's algorithm, and sending matched data to the retailer client. 

To further protect bank's raw data, ARPA's computation network can run retail client's analytical functions directly on matched data and only send back outputs to retail client. Now the bank has full discretion over the usage of its data.

\paragraph{Case 2: Privacy-Preserving Medical Diagnosis}
Individual medical data contains sensitive information that is risky to run diagnosis using 3rd party models or tools. On the other hand, in recent years AI specialists have developed more accurate algorithms to assist medical professional's judgment based on symptoms, medical history, CT scan images, etc. ARPA's privacy-preserving feature allows data to be computed without leaking information to 3rd party model providers. Medical service entities or individuals can also screen for the best diagnosis model on a marketplace, such as genetic analysis. 

\paragraph{Case 3: Identity Authentication}
ARPA is capable of secure identity information storage and authentication for users in dApps and services. Identity information including fingerprint, face, voice unique to user is secret-shared to several nodes. The authentication process validates user identity via a private contract in a completely trustless way. This feature links blockchain with the real world.



\newpage
\section{Roadmap}  \label{sec:roadmap}

\paragraph{Q1 2018} Idea Generation, Data Renting Business Case
\paragraph{Q2 2018} Initial Team Forming \& Funding
\paragraph{Q3 2018} Whitepaper, Tech Notes
\paragraph{Q4 2018} MPC POC Demo, MPC Network Launch
\paragraph{Q1 2019} Testnet 1.0 (ASTRAEA) Release
\paragraph{Q2 2019} Testnet 2.0 (ATLAS) Release

\paragraph{Q3 2019} ARPA Mainnet Release
\paragraph{Q4 2019} Business Case Development, Off-chain Computation Solution
\paragraph{Q1 2020} Security Enhancement, Protocol Improvements


\newpage
\bibliography{main.bib}

\begin{thebibliography}{10}

\bibitem{hyperledger}
Welcome to hyperledger fabric.
\newblock \url{https://www.hyperledger.org/}.

\bibitem{aliasgari2013secure}
{\sc Aliasgari, M., Blanton, M., Zhang, Y., and Steele, A.}
\newblock Secure computation on floating point numbers.
\newblock In {\em NDSS\/} (2013).

\bibitem{atzei2017survey}
{\sc Atzei, N., Bartoletti, M., and Cimoli, T.}
\newblock A survey of attacks on ethereum smart contracts (sok).
\newblock In {\em Principles of Security and Trust}. Springer, 2017,
  pp.~164--186.

\bibitem{bartoletti2017empirical}
{\sc Bartoletti, M., and Pompianu, L.}
\newblock An empirical analysis of smart contracts: platforms, applications,
  and design patterns.
\newblock In {\em International Conference on Financial Cryptography and Data
  Security\/} (2017), Springer, pp.~494--509.

\bibitem{baum2014publicly}
{\sc Baum, C., Damg{\aa}rd, I., and Orlandi, C.}
\newblock Publicly auditable secure multi-party computation.
\newblock In {\em International Conference on Security and Cryptography for
  Networks\/} (2014), Springer, pp.~175--196.

\bibitem{ben2013snarks}
{\sc Ben-Sasson, E., Chiesa, A., Genkin, D., Tromer, E., and Virza, M.}
\newblock Snarks for c: Verifying program executions succinctly and in zero
  knowledge.
\newblock In {\em Advances in Cryptology--CRYPTO 2013}. Springer, 2013,
  pp.~90--108.

\bibitem{bogdanov2008sharemind}
{\sc Bogdanov, D., Laur, S., and Willemson, J.}
\newblock Sharemind: A framework for fast privacy-preserving computations.
\newblock In {\em European Symposium on Research in Computer Security\/}
  (2008), Springer, pp.~192--206.

\bibitem{bogetoft2009secure}
{\sc Bogetoft, P., Christensen, D.~L., Damg{\aa}rd, I., Geisler, M., Jakobsen,
  T., Kr{\o}igaard, M., Nielsen, J.~D., Nielsen, J.~B., Nielsen, K., Pagter,
  J., et~al.}
\newblock Secure multiparty computation goes live.
\newblock In {\em International Conference on Financial Cryptography and Data
  Security\/} (2009), Springer, pp.~325--343.

\bibitem{bonnoron2017somewhat}
{\sc Bonnoron, G., Fontaine, C., Gogniat, G., Herbert, V., Lap{\^o}tre, V.,
  Migliore, V., and Roux-Langlois, A.}
\newblock Somewhat/fully homomorphic encryption: Implementation progresses and
  challenges.
\newblock In {\em International Conference on Codes, Cryptology, and
  Information Security\/} (2017), Springer, pp.~68--82.

\bibitem{randao}
{\sc Buterin, V.}
\newblock Randao beacon exploitability analysis, round 2.
\newblock
  \url{https://ethresear.ch/t/randao-beacon-exploitability-analysis-round-2/1980}.
\newblock Revised on May 12, 2018.

\bibitem{sun2017revisit}
{\sc Chen~Sun, Abhinav~Shrivastava, S. S. A.~G.}
\newblock Revisiting unreasonable effectiveness of data in deep learning era.
\newblock {\em Computer Vision and Pattern Recognition 2\/} (2017).

\bibitem{chor1985verifiable}
{\sc Chor, B., Goldwasser, S., Micali, S., and Awerbuch, B.}
\newblock Verifiable secret sharing and achieving simultaneity in the presence
  of faults.
\newblock In {\em Foundations of Computer Science, 1985., 26th Annual Symposium
  on\/} (1985), IEEE, pp.~383--395.

\bibitem{damgaard2012multiparty}
{\sc Damg{\aa}rd, I., Pastro, V., Smart, N., and Zakarias, S.}
\newblock Multiparty computation from somewhat homomorphic encryption.
\newblock In {\em Advances in Cryptology--CRYPTO 2012}. Springer, 2012,
  pp.~643--662.

\bibitem{ethereum2018}
{\sc Ethereum}.
\newblock Ethereum.
\newblock \url{https://www.ethereum.org/}.
\newblock Accessed Sept 5, 2018.

\bibitem{ethereum2018randnum}
{\sc Ethereum}.
\newblock On-chain random number generation.
\newblock
  \url{https://github.com/ethereum/research/wiki/Problems#21-on-chain-random-number-generation-63}.
\newblock Accessed Sept 5, 2018.

\bibitem{eyal2018majority}
{\sc Eyal, I., and Sirer, E.~G.}
\newblock Majority is not enough: Bitcoin mining is vulnerable.
\newblock {\em Communications of the ACM 61}, 7 (2018), 95--102.

\bibitem{geisler2007viff}
{\sc Geisler, M.}
\newblock Viff: Virtual ideal functionality framework.
\newblock {\em Homepage: http://viff. dk\/} (2007).

\bibitem{truebit}
{\sc Jason~Teutsch, C.~R.}
\newblock A scalable verification solution for blockchains.

\bibitem{plasma}
{\sc Joseph~Poon, V.~B.}
\newblock Plasma: Scalable autonomous smart contracts.

\bibitem{permissioned}
{\sc Kadiyala, A.}
\newblock Nuances between permissionless and permissioned blockchains.
\newblock
  \url{https://medium.com/@akadiyala/nuances-between-permissionless-and-permissioned-blockchains-f5b566f5d483}.
\newblock Revised on Feb 18, 2018.

\bibitem{keller2018overdrive}
{\sc Keller, M., Pastro, V., and Rotaru, D.}
\newblock Overdrive: making spdz great again.
\newblock In {\em Annual International Conference on the Theory and
  Applications of Cryptographic Techniques\/} (2018), Springer, pp.~158--189.

\bibitem{scalability_issue}
{\sc Kumar, U.}
\newblock Understanding ethereum — pertinent problems,scalability, and
  possible solutions.
\newblock
  \url{https://medium.com/coinmonks/understanding-ethereum-pertinent-problems-scalability-and-possible-solutions-eb4fec0405be}.
\newblock Revised on June 2, 2018.

\bibitem{lamport1982byzantine}
{\sc Lamport, L., Shostak, R., and Pease, M.}
\newblock The byzantine generals problem.
\newblock {\em ACM Transactions on Programming Languages and Systems (TOPLAS)
  4}, 3 (1982), 382--401.

\bibitem{banker2016privacy}
{\sc Macheel, T.}
\newblock Banks' privacy concerns shaping blockchain vendors' strategies.
\newblock
  \url{https://www.americanbanker.com/news/banks-privacy-concerns-shaping-blockchain-vendors-strategies}.
\newblock Accessed Sept 5, 2018.

\bibitem{mazieres2015stellar}
{\sc Mazieres, D.}
\newblock The stellar consensus protocol: A federated model for internet-level
  consensus.
\newblock {\em Stellar Development Foundation\/} (2015).

\bibitem{micali2000computationally}
{\sc Micali, S.}
\newblock Computationally sound proofs.
\newblock {\em SIAM Journal on Computing 30}, 4 (2000), 1253--1298.

\bibitem{sgx}
{\sc Miguel~Castro, B.~L.}
\newblock Intel software guard extensions.
\newblock \url{https://software.intel.com/en-us/sgx}.
\newblock Accessed Nov 16, 2018.

\bibitem{pbft}
{\sc Miguel~Castro, B.~L.}
\newblock Practical byzantine fault tolerance.
\newblock \url{http://pmg.csail.mit.edu/papers/osdi99.pdf}.
\newblock Revised on 1999.

\bibitem{nakamoto2008bitcoin}
{\sc Nakamoto, S.}
\newblock Bitcoin: A peer-to-peer electronic cash system.

\bibitem{rationality}
{\sc Palmer, D.~E.}
\newblock Economic rationality.
\newblock \url{https://www.britannica.com/topic/economic-rationality}.
\newblock Revised on Dec 18, 2015.

\bibitem{pedersen1991non}
{\sc Pedersen, T.~P.}
\newblock Non-interactive and information-theoretic secure verifiable secret
  sharing.
\newblock In {\em Annual International Cryptology Conference\/} (1991),
  Springer, pp.~129--140.

\bibitem{rabin2005exchange}
{\sc Rabin, M.~O.}
\newblock How to exchange secrets with oblivious transfer.
\newblock {\em IACR Cryptology ePrint Archive 2005\/} (2005), 187.

\bibitem{rackoff1991non}
{\sc Rackoff, C., and Simon, D.~R.}
\newblock Non-interactive zero-knowledge proof of knowledge and chosen
  ciphertext attack.
\newblock In {\em Annual International Cryptology Conference\/} (1991),
  Springer, pp.~433--444.

\bibitem{schwartz2014ripple}
{\sc Schwartz, D., Youngs, N., Britto, A., et~al.}
\newblock The ripple protocol consensus algorithm.
\newblock {\em Ripple Labs Inc White Paper 5\/} (2014).

\bibitem{shamir1979share}
{\sc Shamir, A.}
\newblock How to share a secret.
\newblock {\em Communications of the ACM 22}, 11 (1979), 612--613.

\bibitem{gaslimit}
{\sc Sharif, A.}
\newblock Ethereum design rationale.
\newblock
  \url{https://github.com/ethereum/wiki/wiki/Design-Rationale#gas-and-fees}.
\newblock Revised on Sept 1, 2018.

\bibitem{sc1997idea}
{\sc Szabo, N.}
\newblock The idea of smart contracts.
\newblock
  \url{http://www.fon.hum.uva.nl/rob/Courses/InformationInSpeech/CDROM/Literature/LOTwinterschool2006/szabo.best.vwh.net/idea.html}.
\newblock Accessed Sept 5, 2018.

\bibitem{turkihoneyledgerbft}
{\sc Turki, H., Salgado, F., and Camacho, J.~M.}
\newblock Honeyledgerbft: Enabling byzantine fault tolerance for the
  hyperledger platform.

\bibitem{GDPR2018}
{\sc Union, E.}
\newblock The eu general data protection regulation (gdpr).
\newblock \url{https://eugdpr.org/}, 2018.
\newblock Enforced May 25, 2018.

\bibitem{randao2016tor}
{\sc vbuterin}.
\newblock Could ethereum do this better? [tor project is working on a web-wide
  random number generator].
\newblock
  \url{https://www.reddit.com/r/ethereum/comments/4mdkku/could_ethereum_do_this_better_tor_project_is/}.
\newblock Accessed Sept 5, 2018.

\bibitem{smartcontract}
{\sc Wikipedia}.
\newblock Smart contract.
\newblock \url{https://en.wikipedia.org/wiki/Smart_contract}.
\newblock Retrived Sep 25, 2018.

\bibitem{yao1982protocols}
{\sc Yao, A.~C.}
\newblock Protocols for secure computations.
\newblock In {\em Foundations of Computer Science, 1982. SFCS'08. 23rd Annual
  Symposium on\/} (1982), IEEE, pp.~160--164.

\end{thebibliography}
\bibliographystyle{acm}
\end{document}